# Gate-imprinted memory and light-induced erasure of superconductivity at $KTaO_3$-based interfaces

Zhihao Chen[1,2,#], Pengxu Ran[3,4,#], Ming Dong Dong[5], Jiexiong Sun[3,4], Fengmiao Li[6,4*], Zhixin Yao[7], Lei Liu[7], Jie Wu[5], Juan Jiang[3,4], Zhi Gang Cheng[1,2,*]

1. Beijing National Laboratory for Condensed Matter Physics and Institute of Physics, Chinese Academy of Sciences, Beijing, China
2. School of Physical Sciences, University of Chinese Academy of Sciences, Beijing, China
3. Hefei National Research Center for Physical Sciences at Microscale, University of Science and Technology of China, Hefei, China
4. School of Emerging Technology, University of Science and Technology of China, Hefei, Anhui, China.
5. Department of Physics, School of Science, Westlake University, Hangzhou, China
6. Hefei National Laboratory, Hefei, Anhui, China.
7. School of Materials Science and Engineering, Peking University, Beijing, China.

#These authors contributed equally.

*Correspondence should be addressed to F.L. (fmli@ustc.edu.cn), and Z.G.C. (zgcheng@iphy.ac.cn)

**Abstract**

Realizing non-volatile control of superconductivity is a key step toward integrating memory and quantum functionality in future information technologies. $KTaO_3$-based heterostructures uniquely host interfacial two-dimensional superconductivity and quantum paraelectric lattice background. The coupling between these two degrees of freedom potentially provides a promising route to encode memory into the superconducting state. Here we reveal two intertwined phenomena in $AlO_x/KTaO_3$ heterostructures: a gate-imprinted memory in which electrostatic gate cycling promotes superconductivity, and its erasure by optical illumination at cryogenic temperatures. These phenomena arise from a previously unrecognized interplay between the superconducting interface and emergent lattice excitations including polar-nanoregion reorientation and charge trapping/detrapping by oxygen vacancies. These results demonstrate configurable superconductivity at correlated oxide interfaces, opening a pathway to enrich dissipationless transport with non-volatile controls for superconducting elements.

## Introduction

Coexistence of electric polarizability and superconductivity[1–5] could potentially offer an exciting research frontier to integrate dissipationless transport with non-volatile functionality. It provides a promising route towards quantum neuromorphic architectures[6–9] which aims to combine memory switch and controllable superconducting coherence for adaptive information processing[10,11]. Integrating these properties within a single material would immensely facilitate the encode of memory into superconducting elements and enable programmable control of quantum states. However, it is unlikely to realize such synergetic functions merely within the framework of two-dimensional electron gases (2DEGs). It instead requires couplings between collective electronic orders and lattice excitations, such as dipolar dynamics, to allow for reversible and reliable modulation.

2DEGs at oxide interfaces[12,13] offer a promising platform to explore this concept because their confined geometry at interfaces amplifies their coupling to the embedding lattice. Emergent phases such as superconductivity[14–19], magnetism[20–23], and topological states[24–29] realized at interfaces of $SrTiO_3$ (STO)- and $KTaO_3$ (KTO)-based heterostructures mostly originate from such coupling. In particular, superconductivity is highly tunable by electrical[18,30] and optical stimuli[31,32], thereby exhibiting exotic phenomena beyond the standard framework of 2DEGs. For instance, electrostatic gating strongly modulates carriers' mobility but only weakly affects their density[17,18]; the superconducting transition temperature is significantly enhanced by gate cycles at cryogenic temperature but can be only recovered by thermal-cycling up to room temperature[30]; and illumination induces non-reciprocal transport by photo-doping[31]. These anomalies suggest an intimate interplay between electronic correlations and lattice dynamics where polarization, interfacial charge distribution, and defect physics play important roles. Nevertheless, these factors are often modulated simultaneously by a single stimulus, which bringing in difficulties in disentangling their individual contributions.

KTO hosts itinerant electrons from Ta's *5d* orbitals with substantially stronger spin–orbit coupling than STO's *3d* electrons[33]. It is thereby proposed as a key ingredient to realize topological superconductivity. While the interfacial 2DEG already benefits from the spatially extended *5d* orbitals, its properties are heavily influenced by defects and polar dynamics. For instance, oxygen vacancies (OVs) play a dual role to donate itinerant electrons and to induce in-gap states as scattering centers[18,23,34]. The defect-driven picture is further enriched by KTO's quantum paraelectricity[35,36] in which quantum fluctuations prevent long-range ferroelectric ordering but allow for incipient ferroelectricity under external electric field and local polar nanoregions (PNRs) stabilized by defects even at zero field[37,38]. The coupling of defects and PNRs to the 2DEG provides a potential basis for superconductivity modulation, yet how such coupling actually affects the superconducting state remains unresolved. Addressing this outstanding question is essential for bridging electron-lattice interactions with functional design principles.

Here we observe two intertwined phenomena for the superconducting 2DEG at $AlO_x$/KTO interface: electrostatic gate cycling imprints persistent memory by enhancing its superconductivity, and optical illumination erases the memory at cryogenic temperatures. By a series of synergetic analyses including dielectric response, relaxation dynamics, and second harmonic generation (SHG), we attribute the memory imprint and erasure to dielectric modulation near the interface via OVs and PNRs. Furthermore, we are able to disentangle their individual roles: OVs induce bistability to the 2DEG via charge trapping/detrapping below ~100 K; PNRs reshape the interfacial potential landscape below ~35 K via their orientation and alignment. Their cooperative actions govern the robustness of superconductivity, thus establishing the basis for non-volatile controls of memristive functionalities in superconducting oxide heterostructures.

**Results**

Sample characterizations

As shown by the device's structure depicted in Fig. 1a, a 2DEG is formed at the $AlO_x$/KTO interface by chemically reducing KTO's (111) surface in a molecular-beam epitaxy (MBE) chamber. The reduction process creates OVs near the KTO surface and donates electrons to the interfacial potential well as illustrated in Fig. 1b. At the same time, the voids of OVs induce in-gap trap states near the interface. Further details of sample preparation are described in Methods. Because of the potential well, charges transfer to the interface and cause band-bending. As shown by X-ray photoelectron spectroscopy (XPS) in Figs. 1c & 1d, characteristic peaks of K $2p$ and $Ta^{5+}$ $4f$ uniformly shift to higher energy by about 0.5 eV with respect to a bare KTO substrate. The shift confirms the down-ward band-bending. In addition, the extra double peaks emerging between 21 eV and 24 eV are ascribed with the reduction from $Ta^{5+}$ to $Ta^{4+}$ which prove the existence of OVs. Three samples (S1, S2, and S3) were fabricated in similar protocols except for different thickness for the $AlO_x$ capping layer. It leads to variations in carrier density and mobility from sample to sample, which are listed in Supplementary Table 1. All samples demonstrate similar phenomena. The results of S1 are reported in the main text of this Article, and results of S2 and S3 are reported in Supplementary Information.

Memory imprinted by gate cycling

The 2DEG was first measured for its superconducting transition under various gate voltages. The gate voltage ($V_g$) was applied to the electrode on the back of KTO substrate and the leakage current to the electrode ($I_g$) was simultaneously monitored. As shown in Figs. 1e & 1f, $V_g$ was first swept from 0 V to -50 V and then returned to 0 V. The superconducting transition clearly shifts to higher temperature in the return sweep instead of retracing its initial path. The enhancement of the transition temperature ($T_c$, defined as sheet resistance $R_s$ drops to 50% of its normal state value) is further confirmed to be irrespective of the initial sweeping direction (see Supplementary Figure 2). In addition, the critical magnetic field ($\mu_0 H_{c2}$) is also enhanced (see Supplementary Figure 5), suggesting the universal promotion of superconducting properties instead of just $T_c$. As $V_g$ cycles further between $\pm 50$ V, $T_c$ continues drifting upward as summarized in Fig. 1g. A similar progressive rise is also observed for $R_s$ at normal state (Fig. 1h) which stems from the decrease in carrier mobility ($\mu$) despite the slight rise in carrier density ($n$) as shown in Supplementary Figure 1. Both $T_c$ and $R_s$ stabilize as closed hysteretic loops after 2 or 3 cycles. The clockwise $T_c$ loop yields a coercive gate bias of 11 V (Supplementary Figure 1). The $R_s$ loop is also clockwise, and its size gradually decays upon warming but remains discernible up to 100 K (see Fig. 1i and Supplementary Figure 6). For comparison, the loops are more obvious for S2 as $V_g$ cycles between $\pm 150$ V as shown in Supplementary Figure 7.

$I_g$ monitors the charge variation to the back-gate electrode as $V_g$ cycles. It consists two possible contributions, including persistent current $(I_p)$ that reflects KTO's conductance, and displacement current ($I_d$) that arises from charge displacement and redistribution. We observe $I_g$ rising to finite values once $V_g$ starts to sweep and always decaying below noise level (< 100 pA) once the sweep is halted at any finite value (see Supplementary Figure 8a-c). It confirms that $I_p \approx 0$ and $I_g$ is predominantly contributed by $I_d$. Interestingly, an asymmetric $I_g$ hysteresis loop is recorded as shown in Supplementary Figure 8a. For $T = 10$ K, a pronounced rise in $I_g$ appears when $V_g$ approaches -50 V in the first cycle, but no symmetric response is seen for the positive cycling branch. It indicates a field-driven irreversible charge exchange between the 2DEG and the embedding lattice near the interface, as verified by the drastic suppression of this hysteresis in subsequent cycles. This statement is further supported by the synchronization $R_s$ increase as the $I_g$ anomaly appears (Supplementary Fig. 8b-f). The asymmetric hysteresis diminishes at higher temperatures. It transforms to be symmetric at 95 K (Supplementary Figure 8g), suggesting that the irreversible charge exchange is overcome by thermal activation. It is notable that the asymmetric loop is also observed for the other two samples (Supplementary Figure 9) as $V_g$ cycles up to $\pm 150$ V. The loop size and onset voltages are correlated with the 2DEGs' carrier density, which further supports the claim of charge exchange taking place near the interface.

Memory erasure by light illumination

To probe the stability of the gate-imprinted memory, we used a laser emitting diode (LED) to illuminate the samples. The LED emits red light of wavelength $\lambda = 630$ nm. The photon energy $h\nu = 1.97$ eV is smaller than the 3.6 eV gap between conduction band minimum (CBM) and valence band maximum (VBM) of KTO[31,39]. The illumination is thus only capable to excite electrons from in-gap defect states to perturb the 2DEG without introducing holes in the valence band. The excited electrons not only raise 2DEG's Fermi level ($E_F$) as shown in Fig. 2a, but also screen the shallow impurities from scattering the 2DEG. A new metastable equilibrium is therefore built once the elevated $E_F$ exerts an opposing potential gradient to prevent further excitations. Figure 2b exhibits $R_s$'s temporal evolution under illumination. Despite its slight drop at 2 K with a reduced illumination power, it was overwhelmed by the significant rise due to heating effect when the LED was turned on, thus preventing reliable analyses at this temperature. However, the heating effect is negligible at 10 K or higher temperatures even the illumination power is 30 times stronger. Instead of abrupt rise, we observed significant drops in $R_s$ at 10 K and 15 K, which suggest emergence of photoconductivity similar as in other oxide superconductors[40,41]. In particular, the photoconductivity was able to persist at 10 K after turning off the LED, making this temperature ideal for reliable evaluation on illumination effects.

Figure 2c plots the $R_s$ variance during a series of illumination/gate-cycle operations. As an illumination period reduces $R_s$ to a baseline of $R_s^0 = 795\ \Omega\cdot\text{sq}^{-1}$, a gate cycle (Type A: $V_g$ cycles as 0 V → $-50$ V → $+50$ V → 0 V) elevates it to $R_s^A = 840\ \Omega\cdot\text{sq}^{-1}$ by an increment of 5.7%. The switch between $R_s^0$ and $R_s^A$ has been reliably repeated 70 times and the retention times for both states last longer than 12 hours (Supplementary Figure 10). Furthermore, these values can be artificially fine-tuned by adjusting illumination period length and gate-cycling amplitude (Supplementary Figure 11). It is interesting to note that the $R_s$ elevation is also sensitive to cycling sequence. Instead of a 5.7% increase for Type A, it is increased by 7.7% to $R_s^B = 856\ \Omega\cdot\text{sq}^{-1}$ when the cycling sequence is reversed (Type B: $V_g$ cycles as 0 V → $+50$ V → $-50$ V → 0 V). Further details revealed by Hall measurements show that $\mu$ and $n$ respectively increase by 4.0% and 1.6% during the illumination (Figs. 2d & 2e). On the other hand, they decrease by similar magnitudes during gate cycles (Figs. 2f & 2g), roughly accounting for the difference between $R_s^0$ and $R_s^A$ or $R_s^B$. We note that $\mu$ drops more in Type B cycle and leads to the larger rise in $R_s^B$. This is consistent with the bistability of the oppositely poled states as shown in Fig. 1i.

Besides modulating the 2DEG's normal state, illumination also resets the memory of $T_c$ enhancement. Figures 2h displays resistance variances by four successive illuminations that lasted for 220 s, 180 s, 290 s, and 9000 s, respectively. The illumination durations are chosen to create stepwise decreases in $R_s$ of roughly equal size from the original state $L_0$ to intermediate states $L_1$, $L_2$, $L_3$, and eventually reaching the final state $L_4$. As shown in Fig. 2i, the superconducting transition is progressively suppressed to lower temperatures. This suppression is able to completely counter-act the $T_c$ enhancement imprinted by gate cycles (Fig. 2j & 2k), demonstrating the reliable switch between the two states.

We further confirm that the illumination is effective for light with other wavelengths. Supplementary Figure 12 compares the illumination effects by red and blue LEDs. The blue light has a shorter wavelength ( $\lambda = 460$ nm) and a larger photon energy ( $h\nu = 2.70$ eV). It correspondingly creates a larger drop in $R_\mathrm{s}$ and suppresses the superconducting transition to lower temperature. This is understood as more electrons are excited from deep in-gap states due to higher photon energy.

Dielectric responses

The memory imprint and erasure strongly indicate a pronounced coupling between the 2DEG and the lattice. To elucidate the underlying mechanism, we investigate the dielectric response of the $AlO_x$/KTO heterostructure in order to probe dynamic excitations in the host lattice. Figures 3a & 3b plot temperature dependence of dielectric constant ($\varepsilon$) and its tangent loss ($\tan\delta$) measured at multiple frequencies. The remarkable increase of $\varepsilon$ upon cooling is attributed to the softening of transverse optical (TO) phonon mode arising from KTO's quantum paraelectric background. Despite its monotonic rise, $\tan\delta$ exhibits two frequency-dependent loss peaks: one near 30 K (denoted as Peak 1) and another near 100 K (denoted as Peak 2). Both peaks exhibit thermally activated behavior and follow the Arrhenius relation: $\omega = \omega_i e^{-E_\mathrm{ai}/k_\mathrm{B}T_\mathrm{pi}}$ where $\omega = 2\pi f$ is the angular frequency of measurement, $T_\mathrm{pi}$ the peak temperature, $E_\mathrm{ai}$ the thermal activation energy, $k_\mathrm{B}$ the Boltzmann's constant, and $\omega_i$ fitting parameters. Here the subscript $i = 1, 2$ denotes the peak index. The Arrhenius plots shown in Fig. 3c extract that $E_\mathrm{a1} = 76$ meV and $E_\mathrm{a2} = 236$ meV.

In previous studies on dielectric properties of bulk KTO, Peak 1 has been observed with similar activation energy and is associated with the excitation of PNRs reorientation[37,38,42,43]. PNRs are induced by local defects such as interstitials, OVs, and antisites[43]. They are expected to induce local spatial-reversal-symmetry breaking which is confirmed to exist in our heterostructure by SHG results (Supplementary Figure 13). PNRs' spatial extension expand at lower temperatures as the SHG signal becomes increasingly prominent below 100 K, in consistence with the softening of TO mode upon cooling[44]. Notably, its trend to saturate below ~30 K coincides with the loss peak at $T_\mathrm{p1}$, suggesting the retarded response of PNRs to external electric field below this temperature.

On the other hand, Peak 2 has not been observed in previous studies on bulk KTO, and it is expected to stem from the $AlO_x$/KTO interface. $E_{a2} = 236$ meV is comparable with the interfacial band-bending of 0.5 eV as extracted by the XPS results (Fig. 1d). It is also close to OVs' ionization energy[45,46]. We therefore reasonably associate it with the interaction between the interfacial 2DEG and shallow OVs via charge trapping. As external electric field cycles, OVs alternate between being neutralized and ionized via effectively trapping and detrapping charges. The trapping becomes more significant as temperature decreases and gives rise to the dielectric loss peak at $T_{p2}$.

Similar dielectric responses are also observed in S2 with $E_{a1}$ and $E_{a2}$ in excellent agreement (Supplementary Figures 14a-c). They evidently suggest the general existence of these two excitations in all $AlO_x$/KTO heterostructures. It is worth to note that both $E_{a1}$ and $E_{a2}$ are much smaller than the photon energy of 1.97 eV. It means the red-light illumination is able to randomize PNRs' orientation and redistribute charges between OVs and 2DEG even at low temperatures.

We further probe $\varepsilon$'s dependence on gate voltage. The sample was first fully illuminated and then gate-cycled at various temperatures while $\varepsilon$ was monitored. Figures 3d-3k show the results at a series of temperatures. $\varepsilon$ is always reduced at large $V_g$, which is consistent with previous studies on bulk KTO[47,48]. However, there are still some distinct features evolving as temperature varies. For instance, $\varepsilon$ converges when $V_g$ returns to 0 V at 150 K irrespective of the sweeping direction (Fig. 3d). In other words, $\varepsilon$ converges for both positive and negative poling states because OVs' charge trapping is ineffective at this temperature due to thermal activation. By contrast, $\varepsilon$ no longer converges below 120 K, and a bistability is developed where $\varepsilon$ for the negative poling state is always smaller (Figs. 3e-3h). When temperature is below 30 K, this bistability is further enhanced by the coercive flipping of PNRs. In addition, the gate sweep induces an irreversible reduction in $\varepsilon$ from the initial state after illumination (Figs. 3i-3k). This is again in agreement with previous studies[47,48] and attributed to the training of randomly oriented PNRs into collective alignment by external electric field. It is interesting to see these distinct lattice dynamics nicely falling into three temperature sections. As exhibited in Fig. 3c, these sections are separated by the trajectories of Peak 1 and Peak 2: 150 K in Section A with $T > T_{p2}$ where charge trapping is ineffective and coherent polar fluctuation is absent; 50~120 K in Section B with $T_{p1} < T < T_{p2}$ where bistability is developed by OVs' charge trapping; and 10~30 K in Section C with $T < T_{p1}$ where the bistability is further promoted by coherent polar fluctuation, and more importantly, irreversible reduction in $\varepsilon$ from its initial state appears due to the training of PNRs' alignment. It is remarkable to note that this reduction can be fully recovered by subsequent illuminations as demonstrated in Figs. 3l-3n. The dielectric recovery underlines all memory imprint and erasure at both resistive and superconducting states as will be discussed later.

The bistability of the lattice is mapped to the interfacial 2DEG. Shallow OVs' Coulomb scattering to the 2DEG becomes stronger when they are ionized at negative poling state whereas it is suppressed when the OVs are neutralized at the positive poling state. The switch between OVs' ionization and neutralization accounts for the clockwise hysteretic loop of $R_{\mathrm{s}}$ as shown in Fig. 1i. Figure 3a compares the size of $R_{\mathrm{s}}$ hysteretic loop ($\Delta_{\mathrm{L}}$) with $\varepsilon$. The loop size is defined as $\Delta_{\mathrm{L}} = (R_{\mathrm{s}}^{-0} - R_{\mathrm{s}}^{+0})/R_{\mathrm{s}}^{\mathrm{ave}}$, $R_{\mathrm{s}}^{+0/-0}$ is sheet resistance of positive/negative poling states as depicted in Fig. 1i, and $R_{\mathrm{s}}^{\mathrm{ave}} = (R_{\mathrm{s}}^{-0} + R_{\mathrm{s}}^{+0})/2$. $\Delta_{\mathrm{L}}$ and $\varepsilon$ share a similar temperature dependence over a wide range extending from 200 K to about 30 K while $\Delta_{\mathrm{L}}$ increases more rapidly at low temperatures. A more careful comparison is presented in Fig. 3o where $\Delta_{\mathrm{L}}$ is linearly dependent on $\varepsilon$ above 35 K and evolves to a steeper rise at lower temperatures. The slope of $\Delta_{\mathrm{L}}$ vs. $\varepsilon$ characterizes how efficiently the lattice bistability affects the 2DEG. With OVs' charge trapping being the sole underlying mechanism above 35 K, the steeper slope below 35 K suggests that the bistability is further enhanced by PNRs. Similar features of $\Delta_{\mathrm{L}}$-$\varepsilon$ dependence is also observed for S2 with 35 K still being the switching point for the two distinct dependences (Supplementary Figure 14d).

It is worth to note that the memory of $T_{\mathrm{c}}$ enhancement can be attributed to the initial irreversible drop of $\varepsilon$. A reduced $\varepsilon$ weakens dielectric screening and narrows the interfacial potential well. It thus favors superconductivity and leads to the enhancement of $T_{\mathrm{c}}$ and $\mu_0 H_{\mathrm{c2}}$[18,30,49]. For the same reason, the smaller $\varepsilon$ for negative poling state (Fig. 3k) can also explain the clockwise loop for $T_{\mathrm{c}}$ in Fig. 1g. Moreover, it is also consistent with the clockwise loop for $R_{\mathrm{s}}$ (Fig. 1h) and the counter-clockwise loop for $\mu$ (Supplementary Figure 1d) because the 2DEG is exposed to the defects more intensively and experiences more scattering at negative poling state.

Thermally activated relaxations

The interfacial band-bending creates potential well to host the 2DEG. In the meantime, it also generates a potential gradient that promotes partial PNR alignment and OV ionization. Although PNRs are randomized and OVs are neutralized by illuminations, both subsystems still tend to relax and comply with this potential gradient with the assistance of thermal activation at elevated temperatures. Likewise, the 2DEG after gate cycles would also relax as aligned PNRs become more disordered and charge distributions reorganize under thermal fluctuations. To probe these relaxation dynamics, we monitor the temporal evolution of $R_s$ after gate cycles and illuminations.

As Fig. 4a demonstrates, $R_\mathrm{s}$ slowly decays after gate cycles and increases after illumination at 35 K. A series of systematic measurements were conducted at multiple temperature between 10 K and 180 K, with the results of post-cycle relaxation presented in Fig. 4b and the results of post-illumination relaxation presented in Fig. 4c. For both types of results, the time-dependence of $R_\mathrm{s}$ can be fitted by $\Delta R_\mathrm{s}(t) = R_\mathrm{s} \cdot (1 - e^{-t/\tau_{1,2}})$, where $\tau_1$ and $\tau_2$ denote the relaxation time constants after gate cycles and illuminations, respectively. Figure 4d exhibits the mutual minima of $\tau_1$ and $\tau_2$ at 35 K and 95 K. Interestingly, these two minima are not only in agreement with the dielectric loss peaks observed in Fig. 3b, but also coincide with the onset of $R_\mathrm{s}$ bistability near 100 K and its deviation from the linear dependence on $\varepsilon$ at 35 K (Fig. 3o). As a result, we conclude that the two minima mark two accelerations of relaxation associated with PNR randomization and charge release from OVs. The fact that $\tau_1$ is in general larger than $\tau_2$ indicates the slower rate for post-cycle state relaxation. It is even barely detectable at intermediate temperatures between 50 K and 75 K (Fig. 4b). This discrepancy reflects the distinct nature of these metastable states. Gate cycling drives the system into a more correlated configuration characterized by aligned PNRs. It is separated from equilibrium by a larger effective energy barrier and therefore relaxes more slowly. In contrast, illumination produces a more disordered and less correlated PNR configuration. It enables faster relaxation through local processes with smaller activation barriers. In addition, the substantial chemical potential difference between shallow and deep in-gap states renders the illuminated state more susceptible to fluctuations, resulting in a shorter $\tau_2$ as photo-excited electrons return to deep in-gap states.

## Discussion

The 2DEG's sensitivities to both electric and optical stimuli suggest its strong coupling to the lattice. This conclusion is supported by three key observations: (1) gate cycles induce pronounced memory and bistable hysteretic effects in both normal and superconducting states; (2) optical illumination effectively erases the memory; and (3) the memory imprint and erasure can be well mapped to dielectric variations. The role of lattice dynamics is further evidenced by dielectric responses which reveal two distinct excitations, OV charge trapping and PNR reorientation, each becoming active below certain temperature scales. In quantum paraelectric lattices such as KTO, these two excitations are often modulated simultaneously by a single stimulus like electrostatic gating, bringing in difficulties to disentangle their individual contributions to the 2DEG. We are able to analyze their respective roles by combining electrostatic gating, optical illumination, dielectric characterization, and nonlinear optical analyses.

Charge trapping is known to induce hysteresis during gate cycling in perovskite oxides which may stem from multiple origins. Studies on STO-based heterostructures claim its origin from electromigration and clustering of OVs[50]. Nevertheless, electromigration and clustering of OVs involve ionic motion which are featured with high activation energies and thus ineffective at ultralow temperatures near 1 K. This is inconsistent with our observation of effective memory imprint at sub-Kelvin temperatures with prompt responses. As a result, these mechanisms are unlikely the dominant factors for our observations in KTO. Furthermore, ferroelastic twin walls have been also claimed as source for charge trapping as their existence in STO originate from the tetragonal-to-cubic structural transition[51]. KTO does not undergo such a transition and is therefore not expected to host a significant density of ferroelastic twin walls. On the other hand, charge trapping may also originate from defect states. For $AlO_x$/KTO heterostructures, it is likely contributed by various types of defects on both sides of the interface. For instance, the amorphous $AlO_x$ capping layer may provide a broad spectrum of trapping energies. Although it may exist, we consistently identify $E_{\mathrm{a2}} = 236$ meV near 100 K across different samples, which strongly points to an alternative and dominant trapping mechanism. Given that $E_{\mathrm{a2}}$ is close to the ionization energy of OVs in the sub-eV range[45,46] and consistent with the interfacial band bending of 0.5 eV on the KTO side, our observations support the primary role of OVs in trapping charges.

In parallel with OVs, PNRs significantly modulate the 2DEG potential well via their internal polar fields. Besides being the main charge donors, OVs are also the major defects to induce

PNRs[43]. Assuming their density about half of carrier density $n$, the volumetric density of PNRs is estimated as $n_{\text{PNR}} = \frac{n}{2t} \approx 5 \times 10^{19}$ cm$^{-3}$, where $n \approx 10^{14}$ cm$^{-2}$ and $t = 10$ nm is the typical thickness of the reduced layer. It gives a PNR concentration $c = a^3 \cdot n_{\text{PNR}} = 0.32\%$ (with lattice constant $a = 3.99$ Å) and an average interfacial polarization $P = P_0 \cdot c = 37$ nC·cm$^{-2}$, where $P_0 = 12$ μC·cm$^{-2}$ is polarization for a supercell containing an OV[43]. The effective internal field $E_{\text{int}} = \frac{P}{\varepsilon\varepsilon_0} = 270$ V·cm$^{-1}$ generates a corresponding potential shift $\delta\phi = E_{\text{int}} \cdot t = 0.27$ meV. By comparison, the external field $E_{\text{ext}} = \frac{V}{t_0} = 1000$ V·cm$^{-1}$ ($V = 50$ V and substrate thickness $t_0 = 0.5$ mm) produces a potential drop $\Delta\phi = 1$ meV across the same region. The ratio $\frac{\delta\phi}{\Delta\phi} \approx 27\%$ closely matches the experimentally observed ratio $\frac{\delta V}{V} = 22\%$ where coercive bias voltage is determined $\delta V = 11$ V from the $T_{\text{c}}$ hysteretic loop as shown in Supplementary Figure 1i.

While PNRs are randomized by optical illumination, the OVs are neutralized by the electrons excited from deep in-gap states. The consequent rises in $n$ and $\mu$ (Figs. 2d & 2e) lead to $R_{\text{s}}$ drop. However, it appears counter-intuitive that $T_{\text{c}}$ is suppressed at the same time, which suggests an underlying mechanism beyond the framework of pure 2D superconductivity. It has been proposed in previous studies that pairing at oxide interfaces is facilitated by soft TO phonon modes[18,49] and coherent polar fluctuations[44,52,53]. The randomized PNRs by illumination renders an increase in $\varepsilon$ and strengthens the screening. It consequently widens the potential well and promotes the 2DEG to penetrating into the KTO lattice. Such an effect is detrimental to the superconductivity and therefore leads to a reduced $T_{\text{c}}$. Reversely, gate cycling reduces $\varepsilon$ and enhances superconductivity.

The counter-actions on both superconducting and normal states provide a proof-of-principle platform for non-volatile functionalities that memory can be electrically written and optically erased. Especially, the optical erasure at cryogenic temperatures provides a complementary control pathway that relaxes the requirement of thermal cycling to room temperature[30]. In the normal state, the resistance variation arises from combined changes in $n$ and $\mu$. Therefore, the resistance contrast can be further enhanced in 2DEGs with lower carrier density. Even in high-density 2DEGs where a given electrostatic modulation produces only a limited change in resistance, the contrast of a few percent can be still precisely identified. The contrast is further tunable by adjusting cycling amplitude. The resistance contrast can be increased from 7.6% for a cycling between ±50 V (Fig. 2c) to 20% for a cycling between ±150 V (Supplementary Figs. 3g and 4j). Such a tunability makes it even possible to extend the operation toward multi-level states. In addition, a $T_c$ variation of around 0.1 K offers an additional degree of freedom to encode information within the superconducting regime where resistance changes are intrinsically amplified near the transition edge. Taken together, our study establishes a physical basis for integrating non-volatile memory with superconducting transport at oxide interfaces. Further studies on device optimization and reproducibility will be important to evaluate their potential for practical implementations.

To summarize, we have demonstrated in this work that the KTO-based 2DEGs exhibits a non-volatile memory in both its superconducting and normal states that is imprinted by electrostatic gate cycles. More significantly, the memory can be erased and reset by optical illumination at cryogenic temperatures. Both operations are realized due to the coupling between the 2DEG and the emergent excitations in the embedding lattice, with PNRs and OVs serving as distinct while cooperative channels for internal gating and Coulomb scattering. Their individual contributions to the non-volatile switching are clearly revealed by the synergetic probes including electrostatic gating, optical illumination, dielectric analyses, and non-linear optical explorations. In a broader context, the coexistence of superconductivity and lattice polarizability exemplifies compelling characters shared with other material families where multiple degrees of freedom cooperate and compete to shape emergent quantum phases. The counter-action afforded by gate cycling and optical illumination offers complementary routes to realize non-volatile and reconfigurable superconducting functionalities. These findings open avenues toward oxide-based superconducting neuromorphic architectures in which quantum coherence and memristive elements are integrated within a unified superconducting platform.

## Methods

Sample preparation

The $AlO_x$/KTO heterostructures were prepared in two steps aiming at creation and stabilization of a 2DEG on the (111) surface of KTO. In the first step, single-crystal KTO substrates with (111) surface and dimensions of 2.5 × 2.5 × 0.5 mm³ were introduced into an MBE chamber and annealed at 500 °C for 1 hour under ultra-high vacuum (pressure lower than $5 \times 10^{-9}$ Torr). In the second step, the substrate temperature was lowered to 350 °C, and a thin layer of aluminum (Al) was deposited. The deposited Al reacted with the oxygen on the KTO surface. This reaction caused oxygen loss and led to the formation of OVs that donated electrons to the 2DEG near the interface. The Al film was further fully oxidized in ambient environment into an $AlO_x$ capping layer. This layer serves as an inert and insulating barrier to protect the reduced KTO surface. It effectively preserved the interfacial 2DEG while shielding it from external perturbations.

SHG measurements

A confocal microscope (WITec, Alpha300RAS) equipped with a 1064 nm excitation laser was used for the SHG measurements. The incident beam passes through a linear polarizer and a half-wave (λ/2) plate before impinging normally on the sample. The SHG signal at 532 nm propagates back through the λ/2 plate and is subsequently analyzed by a second linear analyzer. The sample is mounted onto a holder cooled by a cryostat capable of reaching temperatures as low 4 K.

Electric transport measurements

In order to minimize fringe effect for dielectric measurements, transport measurements were conducted in van der Pauw (vdP) method on the square-shape sample ($2.5 \times 2.5 \times 0.5$ mm$^3$) with four Au wires bonded to the corners. A gold film deposited on the back of the KTO substrate serves as a back-gate electrode. Resistance ($R$) was measured using lock-in amplifier (NF 5640) by applying an ac current excitation of 100 nA at frequency of 13.777 Hz and resolving the voltage amplitude at the same frequency. Sheet resistance ($R_{\mathrm{s}}$) was calculated from $R$ by performing finite-element modelling (FEM). Electrostatic gate voltage ($V_{\mathrm{g}}$) was applied by a source meter (Keithley 2612B) to the back-gate and the leaking current ($I_{\mathrm{g}}$) was monitored by the same source meter.

Dielectric measurements

Dielectric responses were measured by applying an ac voltage excitation $V_{\mathrm{g}}(t) = V_{\mathrm{g}0}e^{i\omega t}$ on the back-gate electrode while keeping the 2DEG shunt to ground. The 2DEG and the back-gate electrode serve as two parallel planar electrodes that sandwich the KTO substrate as dielectric medium. The capacitance of the KTO substrate $C_{\mathrm{KTO}} = \varepsilon_0 \varepsilon \cdot \frac{a_0^2}{t_0}$ varies between 22 and 180 pF as dielectric constant $\varepsilon$ variated between 200 and 1600 across the entire temperature range (Fig. 3a). Here $a_0 = 2.5$ mm and $t_0 = 0.5$ mm are the lateral length and thickness of the KTO substrate, and $\varepsilon_0 = 8.85 \times 10^{-12}$ F/m is permittivity of vacuum. For the measurements with the highest frequency at 4001 Hz, the capacitive impedance of the heterostructure $Z_{\mathrm{KTO}} = \frac{1}{i\omega C_{\mathrm{KTO}}}$ is between 1.80 MΩ and 0.23 MΩ, while its value is even larger for lower frequencies.

The dielectric response was measured using two coaxial cables of about 1 m long and separated by about 2 cm apart. The coaxial shields effectively screen the cables from each other and cut their inductive crosstalk. Their mutual inductance is mainly contributed by the two bare sections of about 1 cm long and 0.1 mm in diameter near the sample holder. These two sections are separated by 1 cm. The effective capacitance is $C = c \cdot l = l \cdot \frac{\pi \varepsilon_0}{\cosh^{-1}\left(\frac{d}{\phi}\right)} = 0.05$ pF, decades smaller than $C_{\mathrm{KTO}}$. Here $d = 1$ cm is the wire separation, $\phi = 0.1$ mm the wire diameter, and $l = 1$ cm the length of bare sections. The mutual inductance of the wirings is estimated as $L = l \cdot \frac{\mu_0}{2\pi} \cdot \cosh^{-1}\left(\frac{d}{\phi}\right) = 10$ nH. Its corresponding inductive impedance at 4001 Hz is $Z_{\mathrm{L}} = \omega L = 0.27$ mΩ, also decades smaller than $Z_{\mathrm{KTO}}$. The wire resistance is 100 Ω each, while the effective resistance between the 2DEG and the back-gate is more than 500 GΩ (leak current $I_{\mathrm{g}} < 10$ pA under 5 V). Therefore, the wire resistance is also negligible.

For the results of temperature dependence shown in Fig. 3a & 3b, the applied sinusoidal voltage is $V_{\mathrm{g0}} = 5$ V in amplitude and its frequency $f = \omega/2\pi$ varies between 0.1 Hz and 4001 Hz. The leakage current measured by lock-in amplifier is given by:

$$I_{\mathrm{g0}} e^{i(\omega t+\theta)} = C \cdot \frac{d\left(V_{\mathrm{g0}} e^{i\omega t}\right)}{dt} = i \frac{\varepsilon \varepsilon_0 A}{d} \cdot V_{\mathrm{g0}} \omega \cdot e^{i\omega t} \tag{1}$$

Taking $\varepsilon = \varepsilon' - i\varepsilon''$, eqn. (1) is converted to

$$\varepsilon'' + i\varepsilon' = \frac{d}{\varepsilon_0 A} \cdot \left(\frac{I_{\mathrm{g0}}}{V_{\mathrm{g0}} \omega}\right) \cdot e^{i\theta} \tag{2}$$

The amplitude of dielectric constant $|\varepsilon| = \frac{d}{\varepsilon_0 A} \cdot \left(\frac{I_{g0}}{V_{g0}\omega}\right)$, and the dielectric tangent loss $\tan\delta = \frac{\varepsilon''}{\varepsilon'} = \cot\theta = \tan\left(\frac{\pi}{2} - \theta\right) \approx \frac{\pi}{2} - \theta$ where the last approximation is valid because the response is mostly capacitive with $\frac{\pi}{2} - \theta \approx 0$.

For the results measured during gate cycles (Figs. 3d-3k), a sinusoidal voltage with amplitude $V_{g0} = 1$ V is superimposed to a sweeping $V_g$ via an isolating capacitor of 2.2 nF. The measurements were conducted at 127 Hz in order to maintain a sufficient signal-to-noise ratio.

Light illumination

Light was emitted by the LED mounted about 1 cm above the sample by applying a dc forward voltage $V_L$. The current $I_L$ was monitored in real time, and the nominal illumination power $P = V_L \cdot I_L$ measures the total output from the current source. There is only a small fraction of the power converted to photons and illuminated on the sample. The fraction is dependent on the position and orientation of the LED with respect to the sample. Less than 5% of the nominal power is estimate to be dissipated on the sample, while it may still cause heating effect and raise sample temperature. For the standard applied illumination power, we estimate that the sample rises by about 0.8 K at 10 K in vacuum. This temperature rise is estimated by the $R_s$ drop when the LED is turned off (see Supplementary Note 9). Given that the $R_s$ drop was only observed in vacuum but disappeared in helium exchange gas, we judge that the heating effect was largely relieved. All the measurements of dielectric responses, $R_s$ hysteretic loops, and relaxations beyond 10 K were conducted with the presence of exchange gas to improve thermal equilibrium.

Monitoring carrier density and mobility during light illuminations and gate cycles

Real-time measurement of the 2DEG's carrier density $n$ and mobility $\mu$ during illuminations and gate cycles is not straight-forward for a vdP setup because $R_s$ needs to be measured under zero magnetic field but $n$ needs to be extracted by Hall measurements under finite magnetic field. We therefore performed three measurements following exactly the same procedures of illuminations and gate cycles, one under zero field and two under magnetic fields with the same magnitude but opposite polarity ($B = \pm 6$ T). We also prepared the initial state for illumination measurement by running the sample through four gate cycles. Similarly, we prepared the initial state for gate cycling measurement by illuminating the sample for sufficiently long period.

## Data availability

The experimental data generated in this study including all figures in the main article and the Supplementary Information have been deposited in the Figshare database[54].

**Acknowledgements**

We appreciate fruitful discussions with Yang Liu, Xi Lin, Qihong Chen, and Ke Zou.

**Funding Statement**

This research is supported by the National Natural Science Foundation of China (Grant No. T2325026 (Z.G.C.)), the National Key R&D Program of China (Grant No. 2021YFA1401902 (Z.G.C.), 2023YFA1406304 (J.J.), 2023YFA1406400 (J.W.), and 2024YFA1408102 (J.W.)), the Key Research Program of Frontier Sciences, CAS (Grant No. ZDBS-LY-SLH0010 (Z.G.C.)), the CAS Project for Young Scientists in Basic Research (Grant No. YSBR047 (Z.G.C.)), and Hefei National Laboratory Self-Deployed Projects (Grant No. ZB2602000301 and ZB2602000302 (F.L.)).

## Author Contributions Statement

Z.G.C and F.L. conceived and designed the project. P.R., J.S., F.L., and J.J. synthesized the heterostructures, Z.C. and Z.G.C. conducted the cryogenic transport measurements. Z.C., Z.Y. and L.L. conducted the dielectric measurements. M.D.D. and J.W. conducted the SHG measurements. F.L. and Z.G.C. wrote the manuscript with inputs from all authors.

## Competing Interests Statement

The authors declare no competing interests.

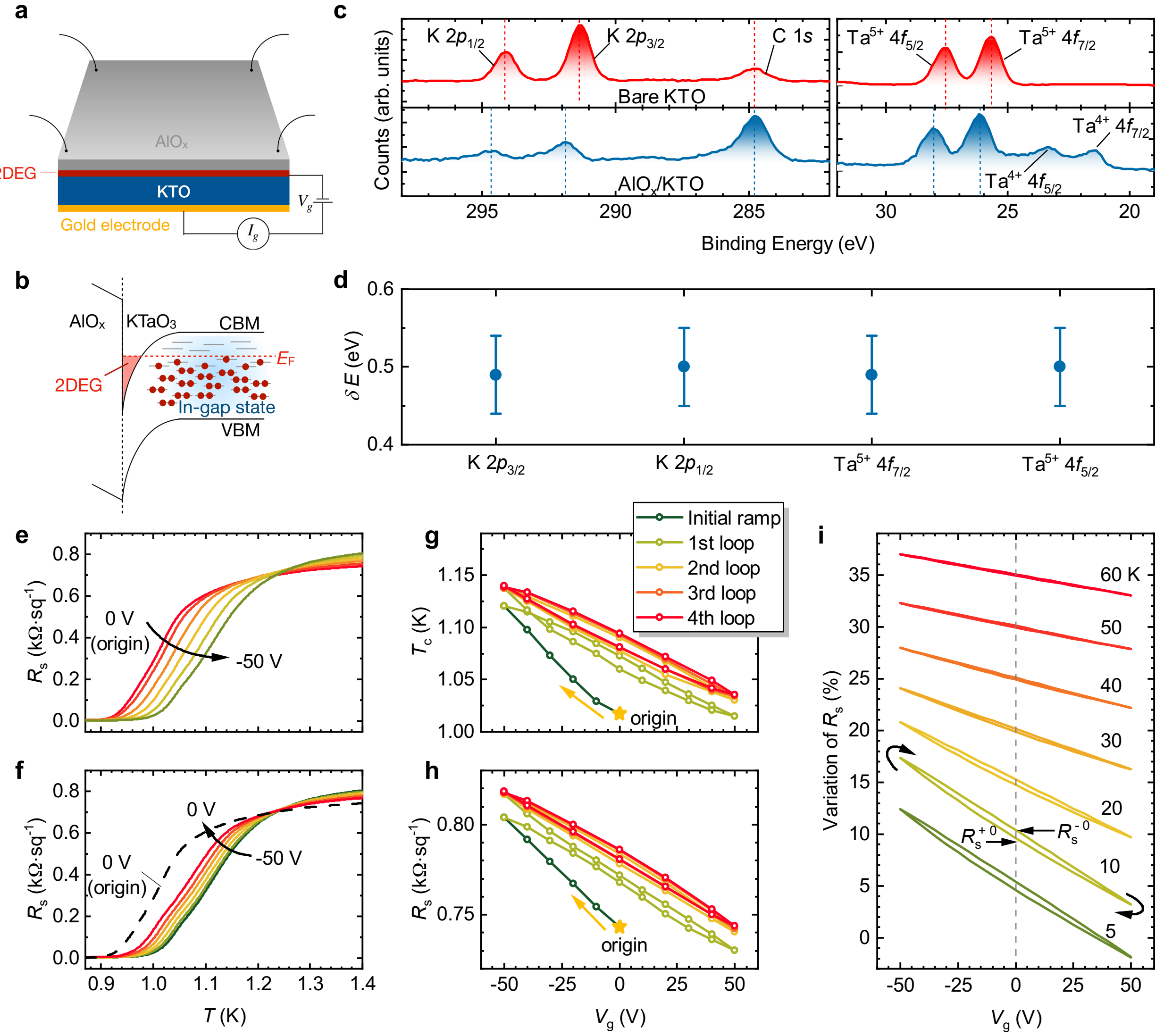

**Figure 1. Memory imprinted by electrostatic gate cycles. a,** Schematic of $AlO_x/KTaO_3$ ($AlO_x$/KTO) heterostructure device. **b,** Band structure near the $AlO_x$/KTO interface. The two-dimensional electron gas (2DEG) is formed at the interfacial potential well caused by band bending. In-gap states between the conduction band minimum (CBM) and the valence band maximum (VBM) are induced by defects like oxygen vacancies (OVs). **c,** X-ray photoelectron spectroscopy (XPS) for a bare KTO substrate (top) and the $AlO_x$/KTO heterostructure (bottom). The device-independent C 1*s* peak arises from $CO_2$ absorption. **d,** Energy shift of the K 2*p* and $Ta^{5+}$ 4*f* peaks for the $AlO_x$/KTO heterostructure with respect to the bare KTO. The error bars of ±0.05 eV are defined by the energy resolution of the XPS measurements. **e & f,** Superconducting transitions measured as gate voltage ($V_g$) cycles in the negative branch at an interval of 10 V. The dashed curve in Panel f represents the original transition measured before the gate cycle. **g & h,** Trajectories of superconducting transition temperature ($T_c$, in Panel g) and sheet resistance at normal state ($R_s$, in Panel h) during a series of gate cycles. $T_c$ is read when sheet resistance $R_s$ drops to half of its normal state value. **i,** Hysteretic loop of $R_\mathrm{s}$ measured at multiple temperatures. Its variation is normalized by the average values $R_\mathrm{s}^\mathrm{ave} = (R_\mathrm{s}^{+0} + R_\mathrm{s}^{-0})/2$, where $R_\mathrm{s}^{+(-)0}$ is sheet resistance for positive (negative) poling state when $V_\mathrm{g}$ crosses 0 V. Data curves are vertically shifted for clarity.

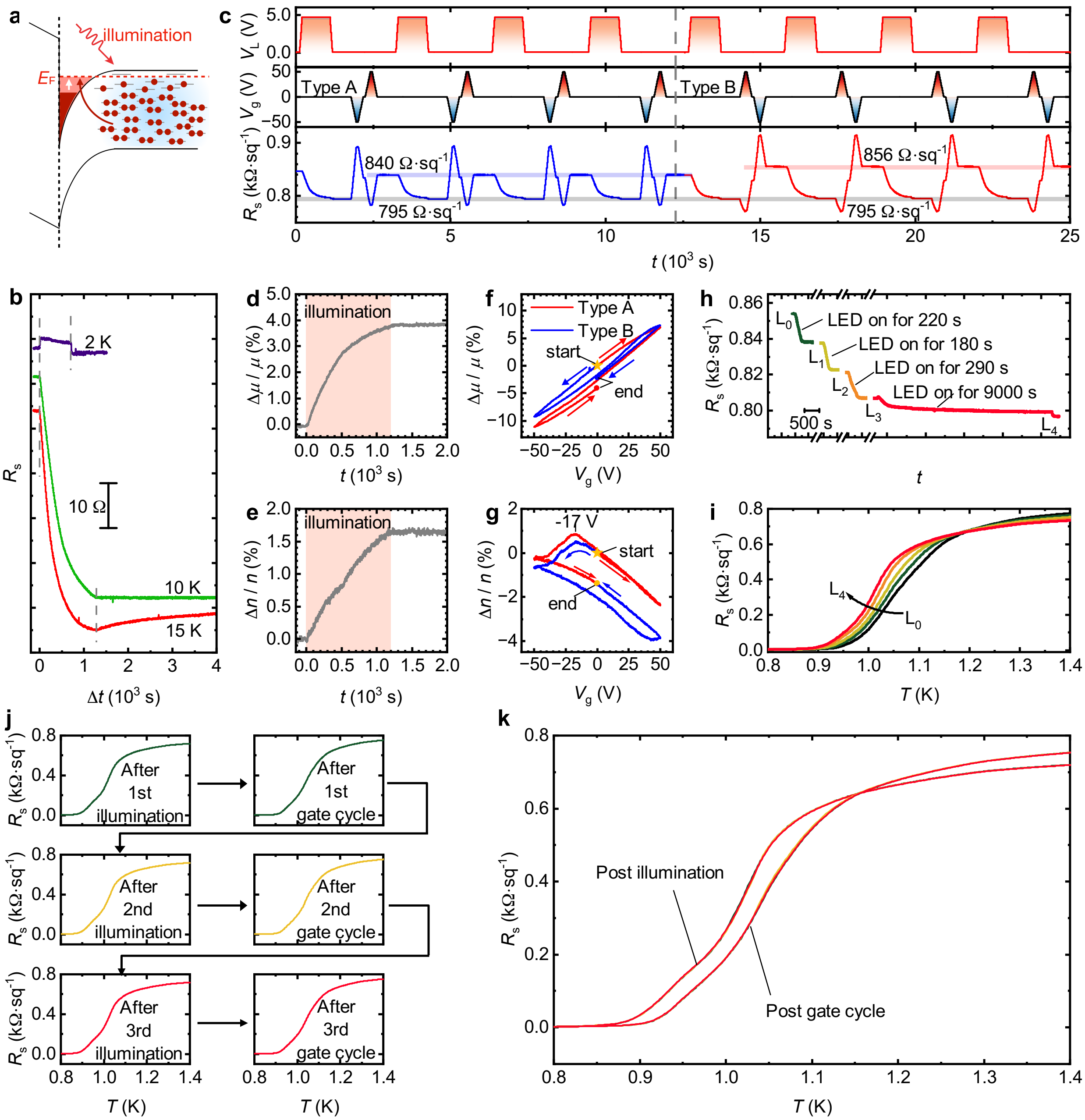

**Figure 2. Memory erasure by optical illumination. a,** Demonstration of illumination. Electrons from deep in-gap states are excited to the 2DEG to raise Fermi level ($E_F$). Shallow OVs are neutralized at the same time. **b,** Resistance variations by illumination. The vertical dashed lines mark the start and end of illumination periods. The nominal illumination power is 23 μW at 2 K and 710 μW at 10 and 15 K. Data curves are vertically shifted for clarity. **c,** Resistance variations during a sequence of illuminations and gate cycles at 10 K. Temporal series of illuminations and gate cycles are shown in the top and middle panels, and the variations of $R_s$ are shown in the bottom panel. The grey bars mark the baselines of $R_s$ after illuminations, and the blue and red bars mark its elevated values after gate cycles. **d & e,** Variations of mobility ($\mu$, in Panel d) and carrier

density ($n$, in Panel e) during the illumination period highlighted by the pink background. **f & g,** Variations of $\mu$ (in Panel f) and $n$ (in Panel g) during gate cycles of Type A (in blue curves) and Type B (in red curves). The beginnings of the cycles are marked by the yellow stars, and the ends are marked by solid circles. **h,** Reductions in $R_s$ during successive illuminations at 10 K. $L_0$ labels the initial state, $L_1 – L_3$ label the intermediate states, and $L_4$ labels the final state. **i,** Superconducting transitions for $L_0 – L_4$. **j,** Superconducting transitions of three rounds of illuminations and gate cycles. The arrows mark the sequence flow. **k,** The same set of data as in Panel j but plotted together. They nicely overlap in two groups, proving the effectiveness of counter-actions on superconducting transition.

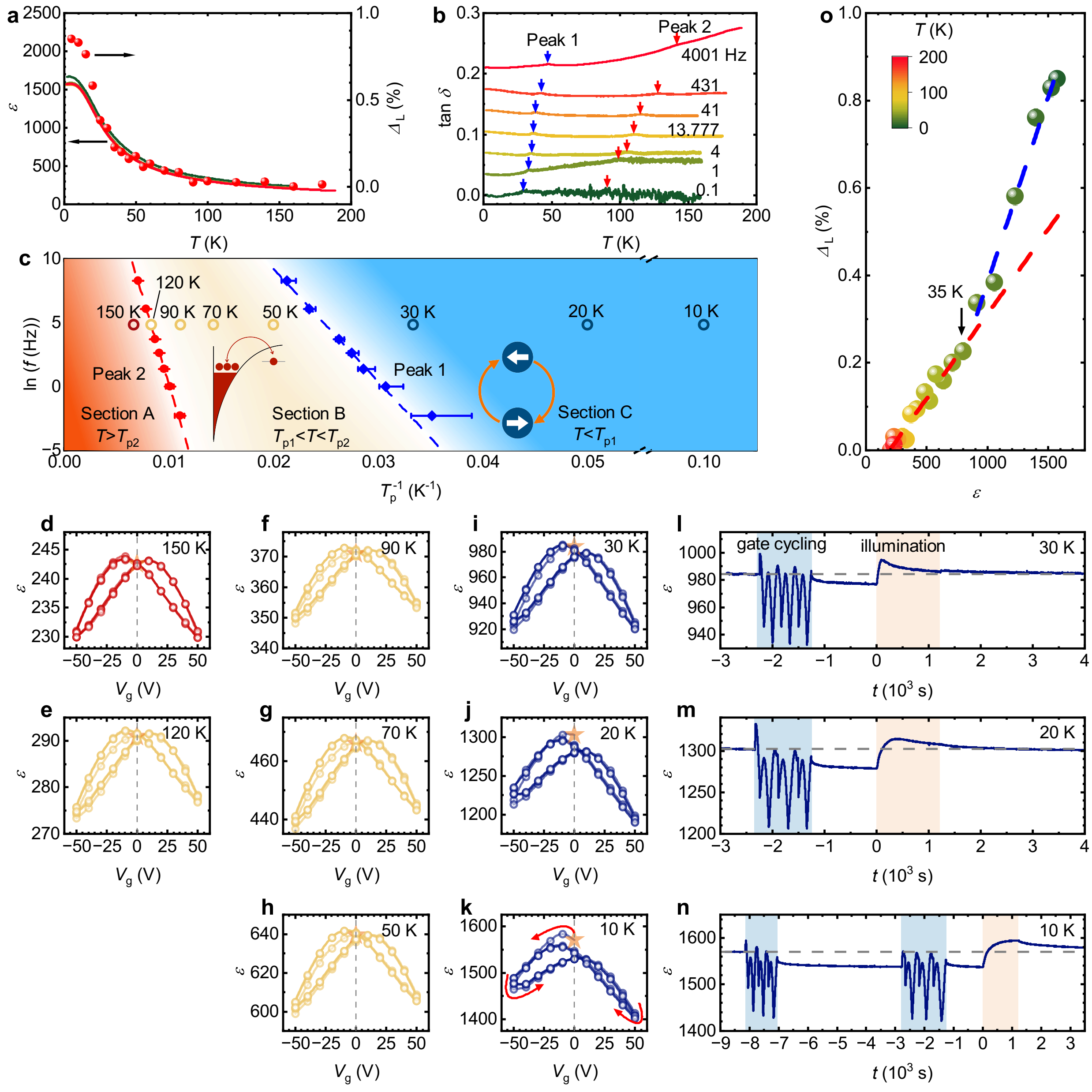

**Figure 3. Dielectric responses of the $AlO_x$/KTO heterostructure. a,** Temperature dependence of dielectric constant ($\varepsilon$) measured at multiple frequencies (plotted by lines) and the size of $R_S$ hysteretic loops ($\Delta_L$, plotted in red circles). **b,** Temperature dependence of dielectric tangent loss ($\tan\delta$) measured at multiple frequencies. Data curves are vertically shifted for clarity with the frequencies of measurements labelled to each curve. Blue and red arrows mark the two dielectric dissipation peaks, respectively. **c,** Arrhenius analyses of the two dissipation peaks. The blue diamonds and red circles represent Peak 1and Peak 2, respectively. The error bars are defined by the temperatures at which the amplitude of dielectric loss drops lower than the peak values by one standard deviation. The inserted cartoons illustrate the underlying physical mechanisms for each excitation. **d-k,** $\varepsilon$ measured during gate cycles at multiple temperatures. The trajectory of gate

cycling is indicated by the red arrows in Panel k. Temperatures of each measurement are tabulated in Panel c (with a fixed frequency of 127 Hz) by open circles. They fall into three sections as highlighted by shadows in different colors. The sections are divided by the fitting lines for $T_{\mathrm{p1}}$ and $T_{\mathrm{p2}}$ as described in the main text. **l-n,** $\varepsilon$ variations as the 2DEG experiences gate cycles (highlighted by the blue shadow) and illuminations (highlighted by the red shadow) at 30 K, 20 K, and 10 K, respectively. The horizontal dashed lines represent the starting values of $\varepsilon$ before gate cycles. It is reduced by gate cycles but recovered by illumination. **o,** Correlation between $\Delta_{\mathrm{L}}$ and $\varepsilon$. $\Delta_{\mathrm{L}}$ is linearly dependent on $\varepsilon$ above 35 K, and deviates to a steeper dependence at lower temperature.

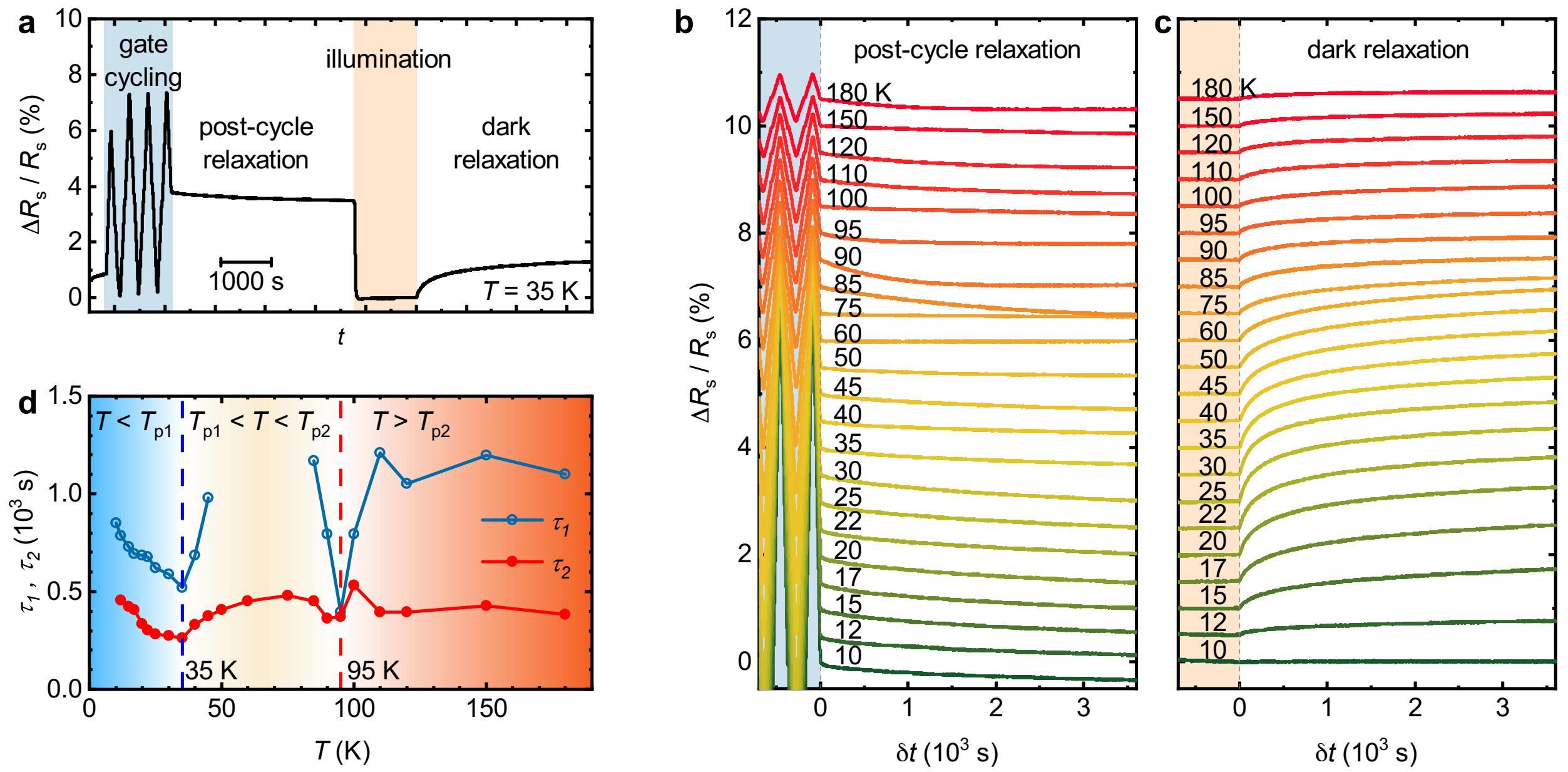

**Figure 4. Relaxations of the 2DEG at various temperatures. a,** Variations of $R_s$ during a series of gate cycles and illumination periods at 35 K. The 2DEG first experienced gate cycles (highlighted by blue shadow) and its relaxation was recorded for one hour. It was then illuminated for 20 minutes (highlighted by red shadow) and another one-hour relaxation was recorded. The variation magnitude is normalized by the value of $R_s$ when the 2DEG was fully illuminated. **b&c,** Variations of $R_s$ for post-cycle relaxation (in Panel b) and dark relaxation (in Panel c) at multiple temperatures. Data curves are vertically shifted for clarity. **d,** Temperature dependence of relaxation time constants $\tau_1$ and $\tau_2$ extracted from data in Panels b & c.

Supplementary Information for

# Gate-imprinted memory and light-induced erasure of superconductivity at $KTaO_3$-based interfaces

Zhihao Chen[1,2,#], Pengxu Ran[3,4,#], Ming Dong Dong[5], Jiexiong Sun[3,4], Fengmiao Li[6,4*], Zhixin Yao[7], Lei Liu[7], Jie Wu[5], Juan Jiang[3,4], Zhi Gang Cheng[1,2,*]

1. Beijing National Laboratory for Condensed Matter Physics and Institute of Physics, Chinese Academy of Sciences, Beijing, China
2. School of Physical Sciences, University of Chinese Academy of Sciences, Beijing, China
3. Hefei National Research Center for Physical Sciences at Microscale, University of Science and Technology of China, Hefei, China
4. School of Emerging Technology, University of Science and Technology of China, Hefei, Anhui, China.
5. Department of Physics, School of Science, Westlake University, Hangzhou, China
6. Hefei National Laboratory, Hefei, Anhui, China.
7. School of Materials Science and Engineering, Peking University, Beijing, China.

#These authors contributed equally.

*Correspondence should be addressed to F.L. (fmli@ustc.edu.cn), and Z.G.C. (zgcheng@iphy.ac.cn)

This file includes

- Supplementary Information Text: Supplementary Note 1 – Supplementary Note 10
- Supplementary Fig. 1 – Supplementary Fig. 15
- Supplementary Table 1

## Supplementary Note 1: Sample specifics

Three $AlO_x$/$KTaO_3$(KTO) heterostructures were fabricated by molecular beam epitaxial (MBE) deposition following the protocols described in Methods. We variated the thickness of Al deposition in order to control the reduction process and the concentration of oxygen vacancies (OVs). It results in a variation in the $AlO_x$ layer thickness. Specifics of the three samples are listed in Supplementary Table 1.

## Supplementary Note 2: Two-dimensional electron gases (2DEGs) during gate cycles

Supplementary Figure 1 demonstrates the trajectories of superconducting transition temperature ($T_c$), sheet resistance ($R_s$) at normal state, carrier density ($n$), and carrier mobility ($\mu$) during gate cycles for S1. Memory effects are observed for all the parameters: $T_c$, $R_s$, $n$ increase, and $\mu$ decreases. In addition, all of them stabilize as closed hysteretic loops after 2~3 cycles. Supplementary Figure 2 presents similar results when $V_g$ is initially swept towards +50 V.

Similar enhancement of $T_c$ is observed for S2 and S3 as $V_g$ cycles between $\pm 150$ V (see Supplementary Figs. 3 & 4). It is notable that $n$ decreases with an increasing $V_g$ for S1. In contrast, $n$ increases with $V_g$ for S2 and S3. Nevertheless, $T_c$ is enhanced for all these samples, suggesting that the memory effect is irrelevant with the dependence of $n$ on $V_g$.

## Supplementary Note 3: Comparison of critical magnetic fields before and after gate cycles

In order to achieve a comprehensive evaluation on superconductivity, its resilience to magnetic field is also studied before and after gate cycles. Magnetoresistances (MRs) of S1 and S2 are measured at various temperatures before and after the gate cycles. The critical field ($\mu_0 H_{c2}$) is determined at 50% of the normal state sheet resistance. Supplementary Figure 5a shows that that $\mu_0 H_{c2}$ is enhanced by the gate cycles at 0.3 K. Supplementary Figure 5b displays that $\mu_0 H_{c2}$ at all temperatures are enhanced. This is in consistence with the $T_c$ enhancement and indicative of a superconductivity strengthening. Similar phenomena are observed for S2 as shown in Supplementary Figures 5c & 5d.

## Supplementary Note 4: Hysteretic loops of sheet resistance at multiple temperatures

Supplementary Figures 6a-6c plots the hysteretic $R_s$ loops at various temperatures. To better demonstrate the loops, Supplementary Figures 6d-6f extract the deviations of the top and bottom branches from their averaged values. It clearly shows a dome-shape dependence on $V_g$ that persists up to 100 K.

Similar hysteretic loops are observed for S2 as $V_g$ cycles between $\pm 150$ V (see Supplementary Fig. 7). The loop sizes are larger than that for S1 because the cycling amplitude extends to higher voltages. Although the loop persists up to 180 K, the loop size no longer shrinks beyond 100 K and remains constant up to 180 K. It may arise from ionic motion of OV impurities driven by large electric field.

**Supplementary Note 5: Hysteretic loop of leakage current to back-gate electrode during gate cycles**

As $V_g$ sweeps, the leakage current flowing to the back-gate electrode ($I_g$) is composed of persistent and displacement current component. We observed that $I_g$ always diminishes to noise level once the sweep is halted at any finite value. The persistent current is therefore negligible, and $I_g$ is mostly contributed by displacement current. It characterizes charge redistribution and polarization within the heterostructure. Supplementary Fig. 8a displays $I_g$ as $V_g$ cycles at 10 K. The first cycle after illumination generates an asymmetric $I_g$ loop. Its magnitude remains negligibly small in the positive branch, but drastically increases beyond -15 V. The increase accelerates as $V_g$ approaches -50 V but decays during to the 10-minute holding time at -50 V. The asymmetry of the loop becomes less obvious in the subsequent three cycles, suggesting an irreversible charge displacement within the sample. This picture is further supported by the series of gate cycles with their amplitudes increased from 5 V to 50 V by a 5 V-step (Supplementary Figs. 8b & 8c). The magnitudes of $I_g$ loops are significantly reduced compared with the direct cycling to $\pm 50$ V. It is remarkable to notice that $I_g$ only starts to rise as $V_g$ sweeps beyond the limit of the previous cycle. Such a phenomenon suggests that the OV distribution is continuous in energy spectrum and their ionization is irreversible. As the cycle expands further, more OVs are ionized and the 2DEG thereby experiences stronger scatterings and becomes more resistive, thus leading to a progressive increase in $R_s$ as shown in Supplementary Fig. 8d. The correlation between the anomalous rises in $I_g$ and $R_s$ is confirmed by Supplementary Figures 8e & 8f. On the other hand, the $I_g$ loop gradually decays as temperature is elevated (Supplementary Fig. 8g,). It also becomes more symmetric near 100 K due to the ineffective charge trapping by OVs.

We further confirm the asymmetric $I_g$ hysteretic loop for other samples. Supplementary Figure 9 compares the hysteretic behaviors for all samples. The rise of $I_g$ firstly appears at -37 V for S2 and boosts to 1 nA at -150 V. This magnitude is smaller than that for S1 due to its lower $n$. On the other hand, $I_g$ starts to rise at -30 V for S3 and reaches 13 nA at -150 V. This onset voltage is between S1 and S2, which is consistent with its intermediate carrier density (see Supplementary Table 1). We thereby conclude that the asymmetric loop of $I_g$ is associated with OVs that play a dual role of contributing itinerant electrons and trapping charges.

**Supplementary Note 6: Counter-action on the memory effect**

Supplementary Figure 10a demonstrates 70 counter-actions between illuminations and gate cycles on the 2DEG at normal state. $R_s$ after each illumination and gate cycle is reproducible. Supplementary Figure 10b demonstrates non-volatility for both states in which $R_s$ does not obviously decay in 12 hours.

Supplementary Figure 11 demonstrates $R_s$ variations in a sequence of six operations, each including an illumination period and a gate cycle. These operations are divided into three groups:

1. In the first group, the illuminations last for 900 s to fully reduce $R_s$ to its baseline value $R_s^0 = 795\ \Omega\cdot\text{sq}^{-1}$. The gate cycles sweep between ±40 V and raise $R_s$ to its elevated value $R_s^a = 843\ \Omega\cdot\text{sq}^{-1}$.

2. In the second group, the illumination periods are kept at 900 s but the gate cycles expand to ±50 V, leading to a higher sheet resistance $R_s^b = 856\ \Omega\cdot\text{sq}^{-1}$. Nevertheless, the same baseline level can be recovered by the illuminations.

3. In the last group, the illumination periods are shortened to 300 s. They are only able to reduce $R_s$ to 808 $\Omega\cdot\text{sq}^{-1}$. However, the gate cycles between ±50 V still raise it to $R_s^b = 856\ \Omega\cdot\text{sq}^{-1}$ despite its higher baseline.

Results of this cyclic protocol comprehensively demonstrates the non-volatility for both the baseline and elevation of $R_s$. They can be reliably adjusted via multiple tuning knobs, including illumination period, illumination fluence, and gate cycling amplitude.

Supplementary Figure 12 demonstrates the effects of illuminations by red and blue light. Blue light generates a larger drop in $R_s$ and a more obvious $T_c$ suppression although its nominal illumination power is only 4.5% of the red light (30 μW vs. 660 μW). It is likely because more electrons are excited from deeper in-gap states to the 2DEG due to the more energetic photons, thereby leading to larger contrasts in $R_s$.

**Supplementary Note 7: Second harmonic generation (SHG) results**

Polar nanoregions (PNRs) can be induced to the quantum paraelectric background of KTO's lattice by interstitials, impurities, antisites, and local strains. The spatial range for their coherent polar fluctuation is given by $l_0 = v_{TO}/\omega_{TO}$, where $v_{TO}$ and $\omega_{TO}$ are group velocity and angular frequency of transverse optical (TO) phonon mode, respectively. As a results, PNRs expand in space as the TO mode softens ($\omega_{TO}$ decreases) during cooling. They thereby play a more significant role at lower temperatures in introducing local symmetry-breaking and polarization. To confirm their existence in our $AlO_x$/KTO heterostructure, we performed SHG measurements to detect symmetry-breaking associated with polarization. Supplementary Figure 13 plots the SHG intensity as a function of temperature. The intensity starts to emerge above noise level below 100 K and rises upon further cooling. In addition, the $C_6$-symmetric angular distribution of the SHG intensity transforms to be $C_2$-symmetric when temperature cools from 300 K to 4 K. The SHG results confirm PNRs' appearance and spatial expansion at low temperatures. Furthermore, the intensity tends to saturate below 40 K, implying that dynamics of PNRs' reorientation becomes slower at in this temperature range.

**Supplementary Note 8: Dielectric responses of S2**

Supplementary Figures 14a & 14b present the temperature dependence of dielectric constant ($\varepsilon$) and tangent loss ($\tan\delta$) for S2. Two loss peaks can be still resolved. The Arrhenius analysesshown in Supplementary Figure 14c extract $E_{a1} = 79$ meV and $E_{a2} = 235$ meV, in excellent agreement with the results for S1.

The correlation between size of hysteretic loop ($\Delta_{\mathrm{L}}$) and $\varepsilon$ is shown in Supplementary Figure 14d. Similar as for S1, the linear dependence below 100 K and the deviation below 35 K are also observed. It is notable that $\Delta_{\mathrm{L}}$ remains finite and temperature-independent above 100 K. It could be due to ionic motion driven by the large bias voltage up to 150 V. This behavior is not detected for S1 with gate cycles limited within $\pm 50$ V.

**Supplementary Note 9: Bolometric effect and thermal equilibrium**

Illumination at cryogenic temperatures may lead to heating effect and cause the rise of sample's temperature from the thermal bath. In our study, the typical nominal illumination power on the LED is 710 μW. Taking a flux conversion rate of $\eta = 20\%$ by presumption, the power for actual illuminating fluence is about 150 μW. Furthermore, we intentionally placed the sample out of LED's viewing angle (40°) in order to minimize heating effect. Therefore, the majority of energy dissipation is not on the sample but on the cold finger and radiation shield of the fridge. Rise of sample's temperature is thereby minimized.

We evaluate the heating effect based on the resistance variation. As shown in Supplementary Fig. 15a, there is an extra drop in $R_{\mathrm{s}}$ by 2.5 $\Omega \cdot \mathrm{sq}^{-1}$ at the end of an illumination period when the LED is turned off. Such a drop is caused by the sample's equilibrium to the thermal bath. It suggests that the sample's temperature is higher than the thermal bath under illumination. According to the $R_{\mathrm{s}} - T$ dependence, $dR_{\mathrm{s}}/dT = 3$ $\Omega \cdot \mathrm{sq}^{-1} \cdot \mathrm{T}^{-1}$ at 10 K. The drop of 2.5 $\Omega \cdot \mathrm{sq}^{-1}$ corresponds to a temperature elevation of 0.8 K. To the sharp contrast, such a drop is not observed when the sample space is filled with helium exchange gas. The disappearance of such a resistance drop indicates the sample in good thermal equilibrium with the thermal bath with the help of exchange gas. All the measurements above 10 K were conducted with the presence of exchange gas, while all the measurements below 10 K were conducted under vacuum and without continuous illumination.

We further illuminated the sample using different fluences. Supplementary Figure 15c compares the resistance variation as the sample is illuminated by nominal power of 1090 μW vs. 400 μW. The same amount of $R_{\mathrm{s}}$ reduction is achieved although the variation rate is higher for larger power. It therefore excludes heating effect as reason for the $R_{\mathrm{s}}$ reduction.

**Supplementary Note 10: Variations of $n$ and $\mu$ during illuminations and gate cycles**

Figs. 2d & 2e exhibit 4% increase in $\mu$ and 1.6% increase in $n$ under illumination. Subsequent gate cycling counteracts this effect: $n$ decreases by 1.39%, nearly restoring its pre-illumination value, and $\mu$ decreases by 4.14% in the Type B cycle, also returning close to the initial state. This correspondence indicates that the mobility change is primarily governed by Coulomb scattering from ionized impurities. The smaller drop of 2.18% in the Type A cycle reflects the hysteretic nature of the dielectric environment.

To quantify this effect, we adopt a charged-impurity-limited Drude picture. In a 2DEG, the mobility can be expressed as $\mu \propto \frac{\alpha(\varepsilon)}{N_{\mathrm{imp}}^{+}}$, where $N_{\mathrm{imp}}^{+}$ is the density of ionized impurities and $\alpha$ is an effective parameter characterizing the sensitivity of the 2DEG to impurities. Assuming that both the dominant changes in carrier density $n$ and $N_{\mathrm{imp}}^{+}$ arise from the photo-excited electrons from the deep in-gap states, we approximate $\frac{\delta N_{\mathrm{imp}}^{+}}{N_{\mathrm{imp}}^{+}} = -\frac{\delta n}{n}$, which yields $\frac{\delta\mu}{\mu} = \alpha\frac{\delta n}{n}$.

According to experimental results in Figs. 3f & 3g, $\frac{\delta n}{n} = -1.39\%$, $\frac{\delta\mu_{\mathrm{A}}}{\mu_{\mathrm{A}}} = -2.18\%$, $\frac{\delta\mu_{\mathrm{B}}}{\mu_{\mathrm{B}}} = -4.14\%$. Therefore, $\alpha_{\mathrm{A}} = \frac{\delta\mu_{\mathrm{A}}}{\mu_{\mathrm{A}}} / \frac{\delta n}{n} \approx 1.57$ for Type A cycles and $\alpha_{\mathrm{B}} = \frac{\delta\mu_{\mathrm{B}}}{\mu_{\mathrm{B}}} / \frac{\delta n}{n} \approx 2.98$ for Type B cycles. The difference between $\alpha_{\mathrm{A}}$ and $\alpha_{\mathrm{B}}$ reflects the secondary modulation from the dielectric screening. We note that the bistability of $\varepsilon$ at low temperature arises from both the charge state of OVs and the orientation of PNRs. The positive poling state gives rise to a larger $\varepsilon$ (see Figs. 3e – 3k). It thereby leads to more effective screening, making the 2DEG less sensitive to the scattering sites and, as a result, a smaller $\alpha$ at the end of Type A cycles. To the contrary, the negative poling state at the end of Type B cycles renders a smaller $\varepsilon$, resulting in weaker screening. It leaves the 2DEG more sensitive to the impurities and therefore a larger $\alpha$.

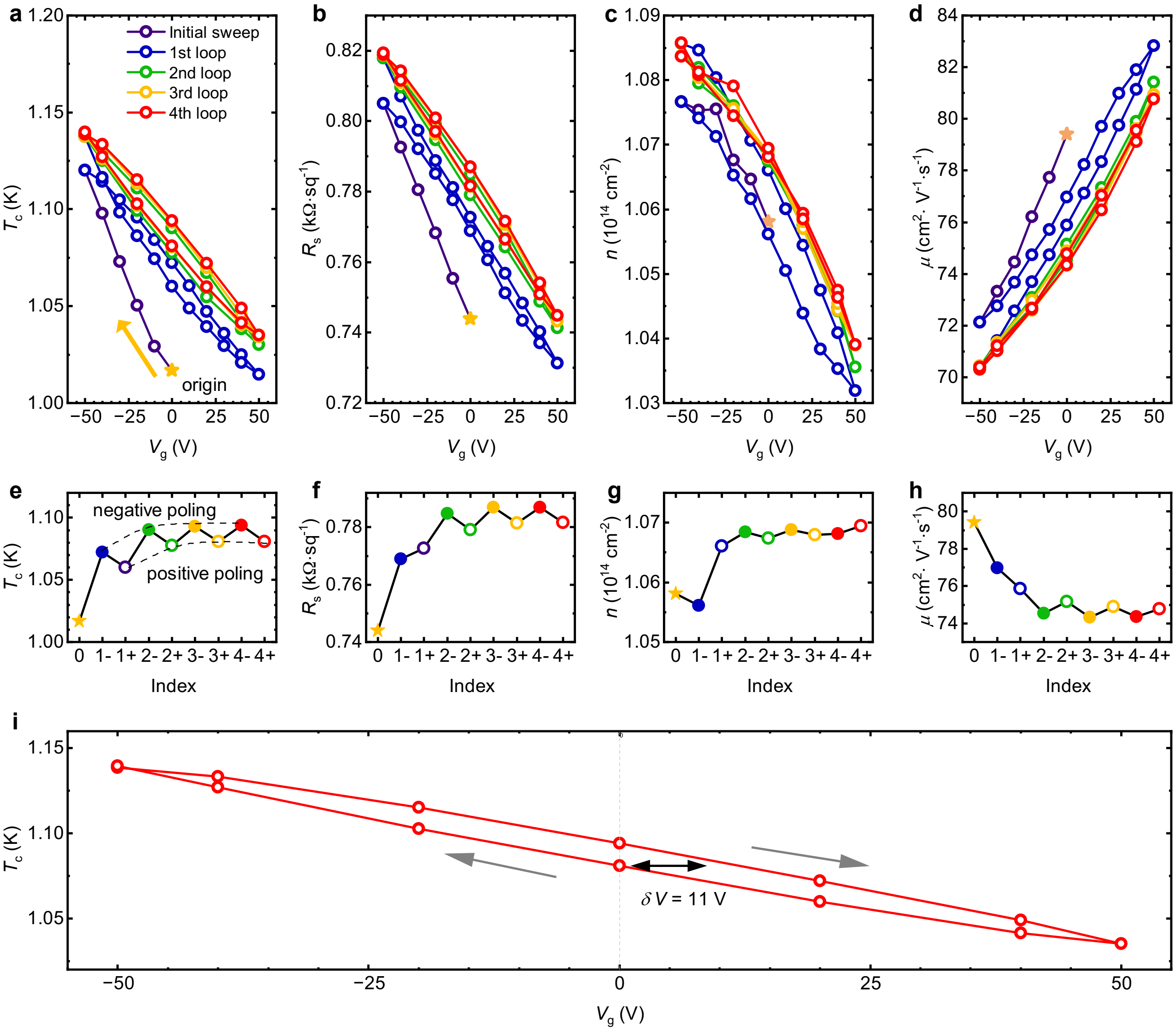

**Supplementary Figure 1. Evolutions of the two-dimensional electron gas (2DEG) for S1 during gate cycles. a-d,** Trajectories of superconducting transition ($T_c$), sheet resistance ($R_s$), areal electron density ($n$), and electrons' mobility ($\mu$) as gate voltage ($V_g$) cycles. The initial state is marked by yellow stars. **e-h,** Evolutions of these parameters at positive and negative poling states when $V_g$ crosses 0 V. The horizontal axes label the cycle index and sweeping direction. For example, 1- denotes the negative poling state in the first cycle. Their color code follows that in the legends of Panel a. **i,** $T_c$ hysteresis in the 4th cycle as plotted in Panel a. A coercive bias gate voltage $\delta V$ is estimated to be 11 V.

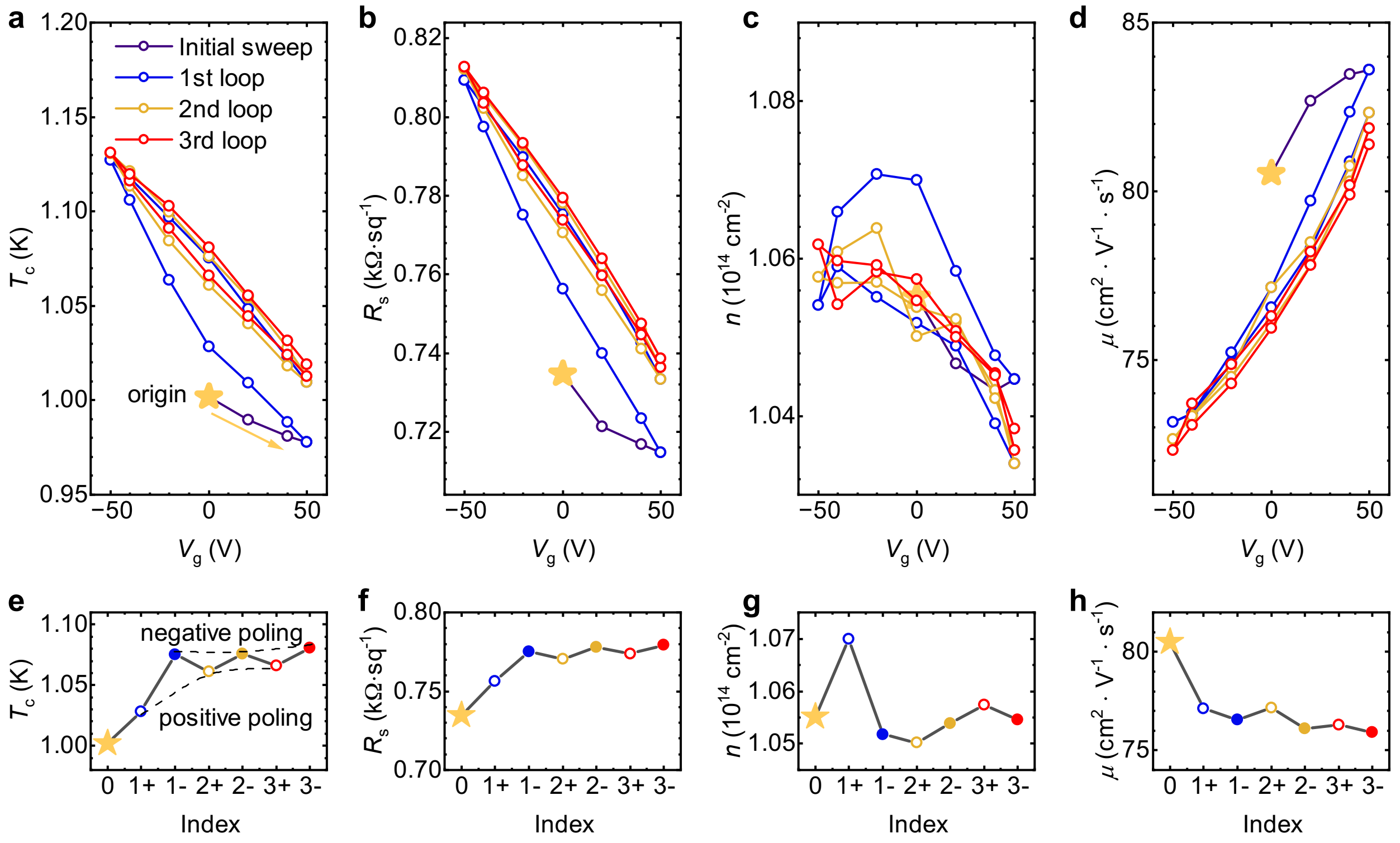

**Supplementary Figure 2. Evolutions of the 2DEG for S1 during gate cycles with the initial sweep towards +50 V. a-d,** Trajectories of $T_c$, $R_s$, $n$, and $\mu$. **e-h,** Evolutions of these parameters at positive and negative poling states. The initial state is marked by yellow stars.

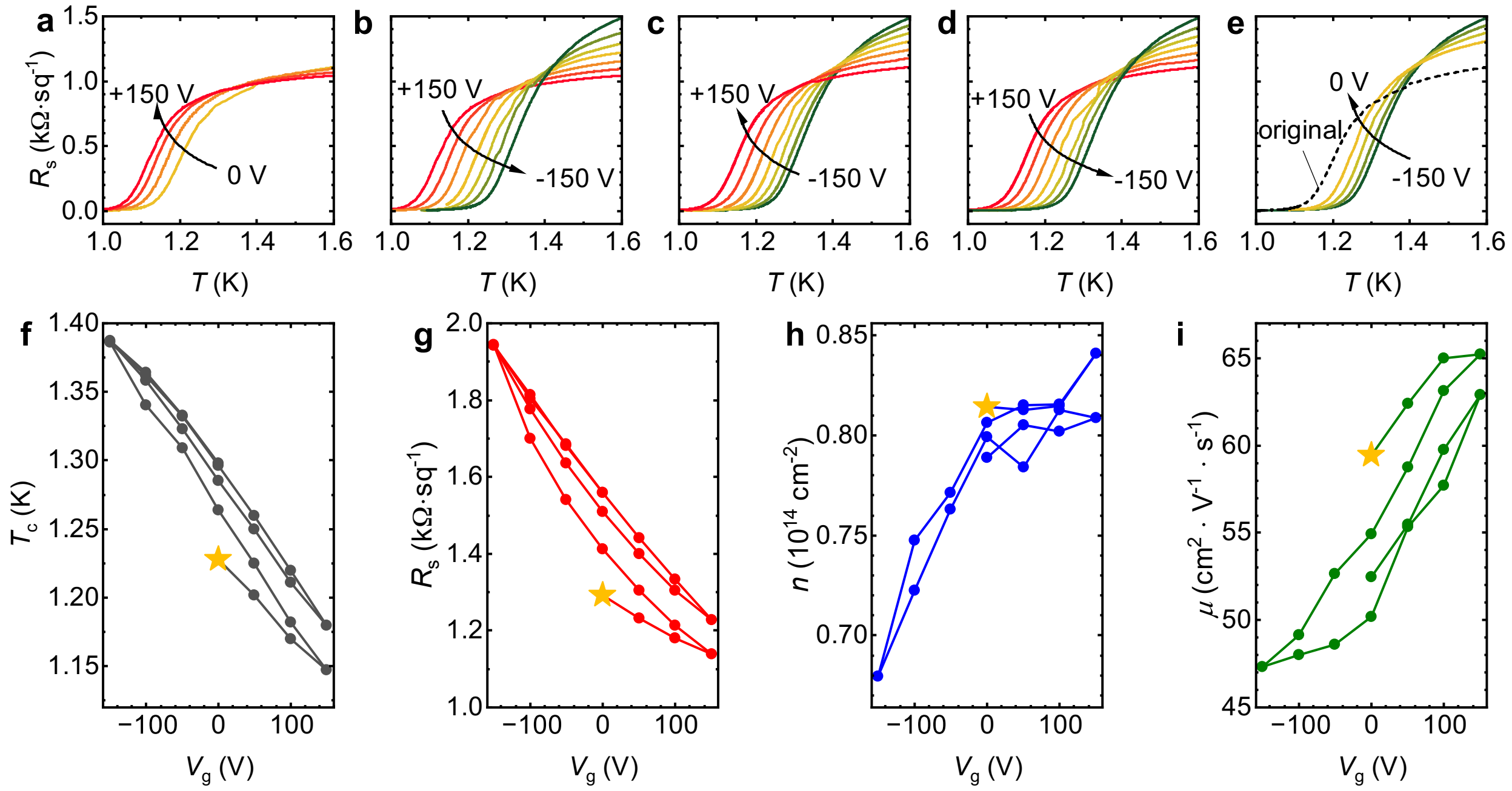

**Supplementary Figure 3. Evolutions of the 2DEG for S2 during gate cycles. a-e,** Superconducting transitions as $V_g$ cycles between $\pm 150$ V by a step interval of 50 V. **f-i,** Trajectories of $T_c$, $R_s$, $n$, and $\mu$ during the gate cycles. The initial state is marked by yellow stars.

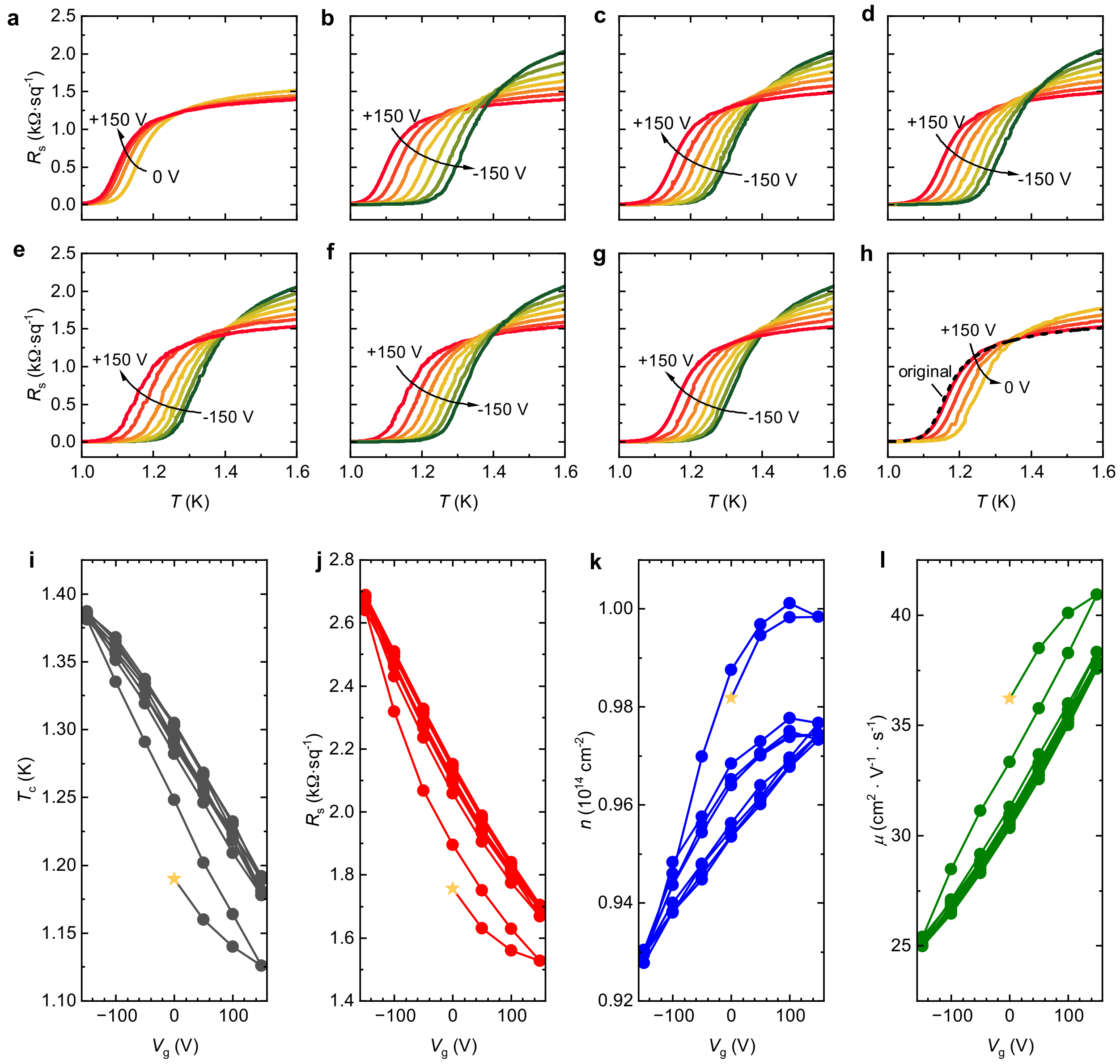

**Supplementary Figure 4. Evolutions of the 2DEG for S3 during gate cycles. a-h,** Superconducting transitions as $V_g$ cycles between $\pm 150$ V by a step interval of 50 V. **i-l,** Trajectories of $T_c$, $R_s$, $n$, and $\mu$ during the gate cycles. The initial state is marked by yellow stars.

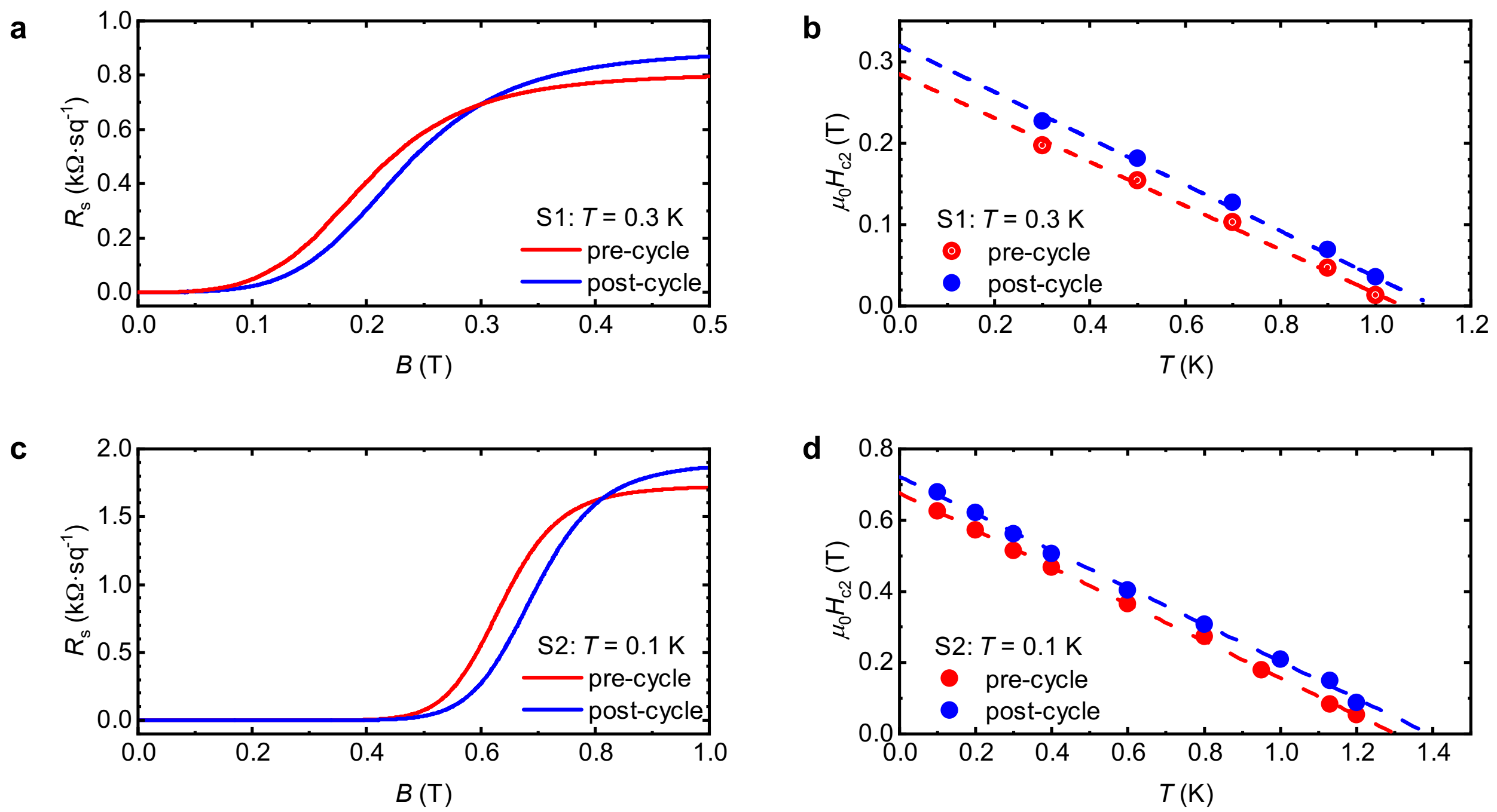

**Supplementary Figure 5. Critical fields ($\mu_0 H_{c2}$) before and after gate cycles. a,** Magnetoresistances (MRs) measured at 0.3 K for S1 before and after gate cycles. **b,** Temperature dependence of $\mu_0 H_{c2}$ for pre- and post-cycle states of S1. **c,** MRs measured at 0.1 K for S2 before and after gate cycles. **d,** Temperature dependence of $\mu_0 H_{c2}$ for pre- and post-cycle states of S2.

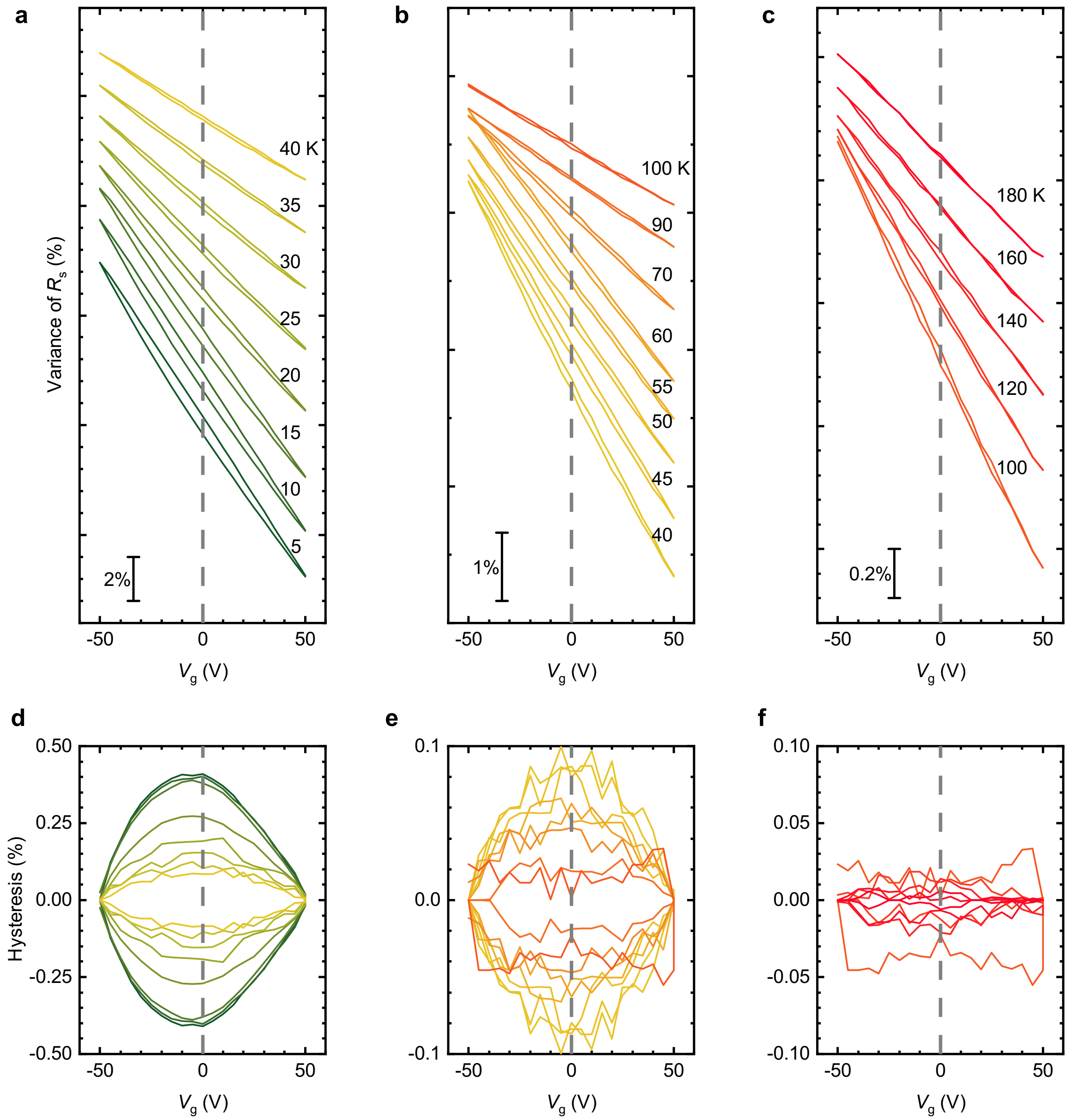

**Supplementary Figure 6. Full data set of hysteretic loops for S1. a,** $R_s$ loops from 5 K to 40 K. **b,** $R_s$ loops from 40 K to 100 K. **c,** $R_s$ loops from 100 K to 180 K. Data are vertically shifted for clarity and the temperatures of measurements are labelled next to each curve. **d-f,** Hysteretic deviation of the top and bottom branches from their averaged values. The curves share the same color code as in Panels a-c.

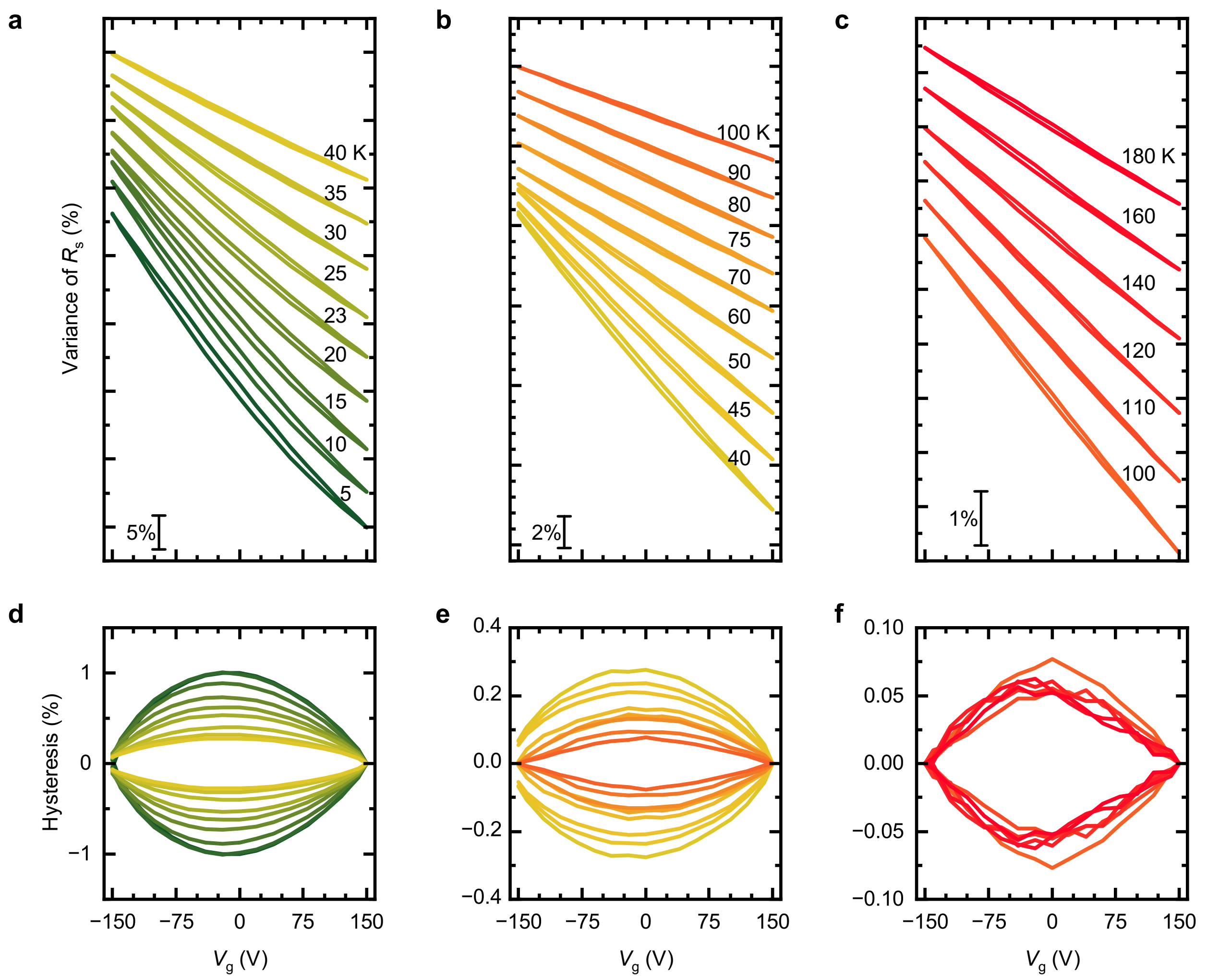

**Supplementary Figure 7. Full data set of hysteretic loops for S2. a,** $R_s$ loops from 5 K to 40 K. **b,** $R_s$ loops from 40 K to 100 K. **c,** $R_s$ loops from 100 K to 180 K. Data are vertically shifted for clarity and the temperatures of measurements are labelled next to each curve. **d-f,** Hysteretic deviation of the top and bottom branches from their averaged values. The curves share the same color code as in Panels a-c.

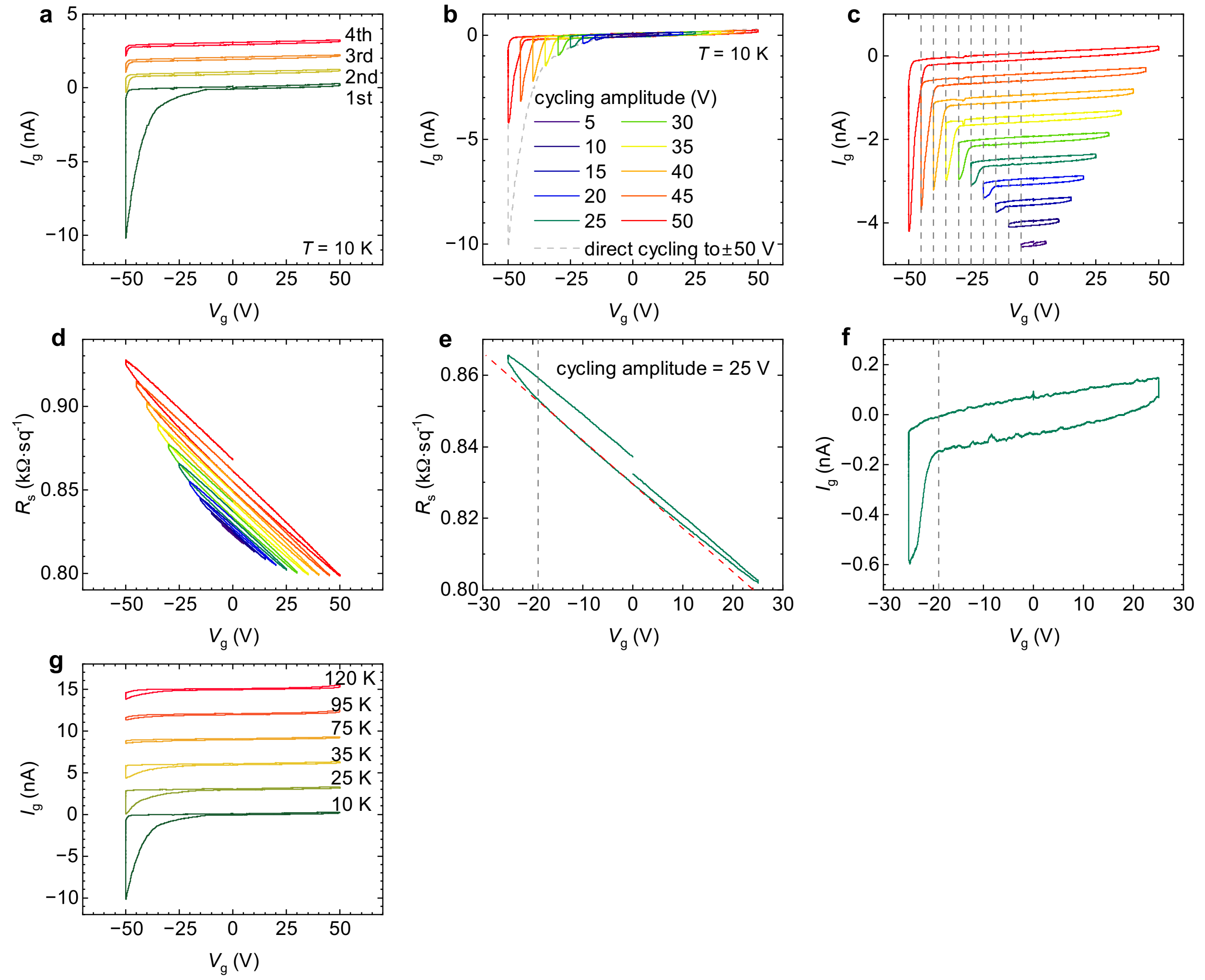

**Supplementary Figure 8. Leakage current to back-gate electrode ($I_g$) monitored during gate cycles. a,** $I_g$ measured in four consecutive sweeps at 10 K. **b,** $I_g$ measured during a series of cycles as the cycling amplitude increases from 5 V to 50 V at a step of 5 V. The dashed curve represents the $I_g$ loop measured when $V_g$ directly cycles between ±50 V as a reference. **c,** The same data in Panel b but vertically shifted. The vertical dashed lines mark the negative extrema of the cycles. **d,** $R_s$ monitored during the cycle series. **e&f,** Correlation between $R_s$ and $I_g$ during the cycle up to ±25 V. The vertical dashed line at -19 V marks the simultaneous abnormal rises of $I_g$ and $R_s$. **g,** $I_g$ loops measured at multiple temperatures.

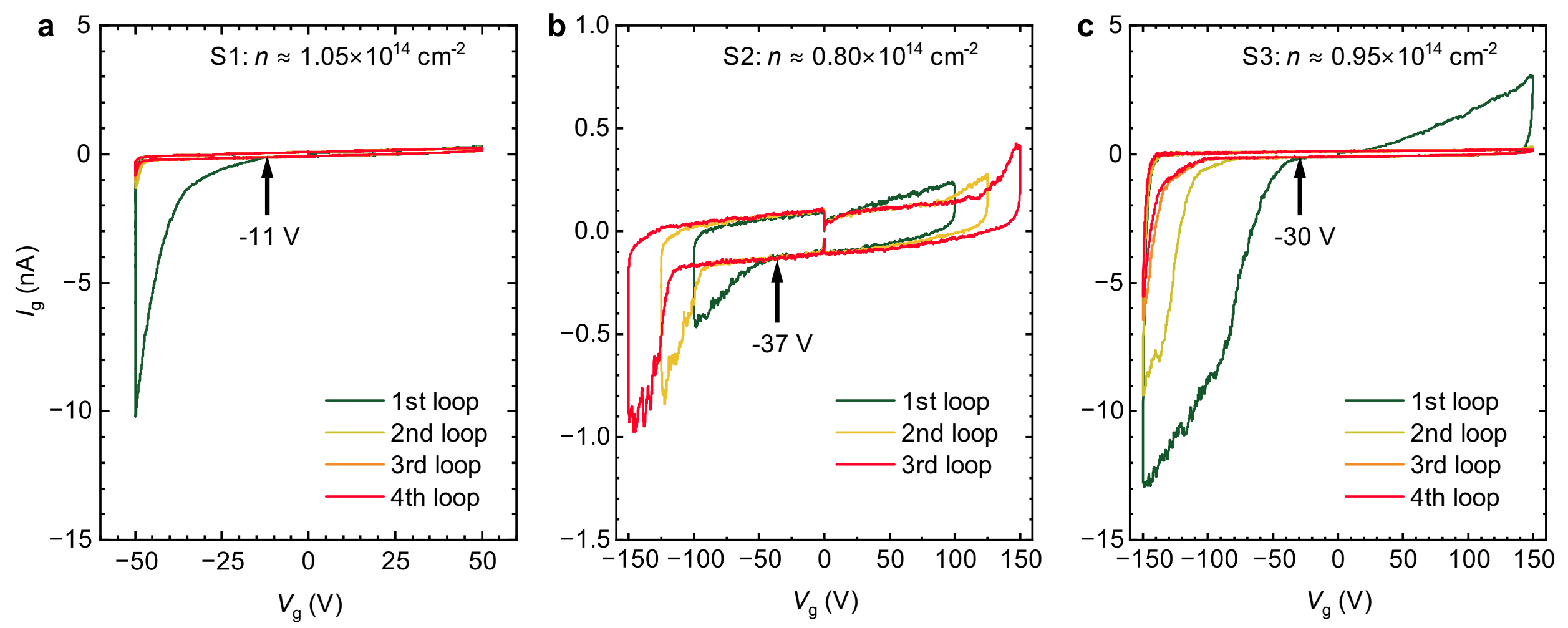

**Supplementary Figure 9. $I_g$ loops for all three samples. a,** $I_g$ monitored for S1 during four gate cycles between ±50 V. It is the same set of data as in Supplementary Fig. 8a. **b,** $I_g$ monitored for S2 during three gate cycles. The cycling amplitude increases from 100 V to 150 V at an interval of 25 V. **c,** $I_g$ monitored for S3 during four cycles between ±150 V.

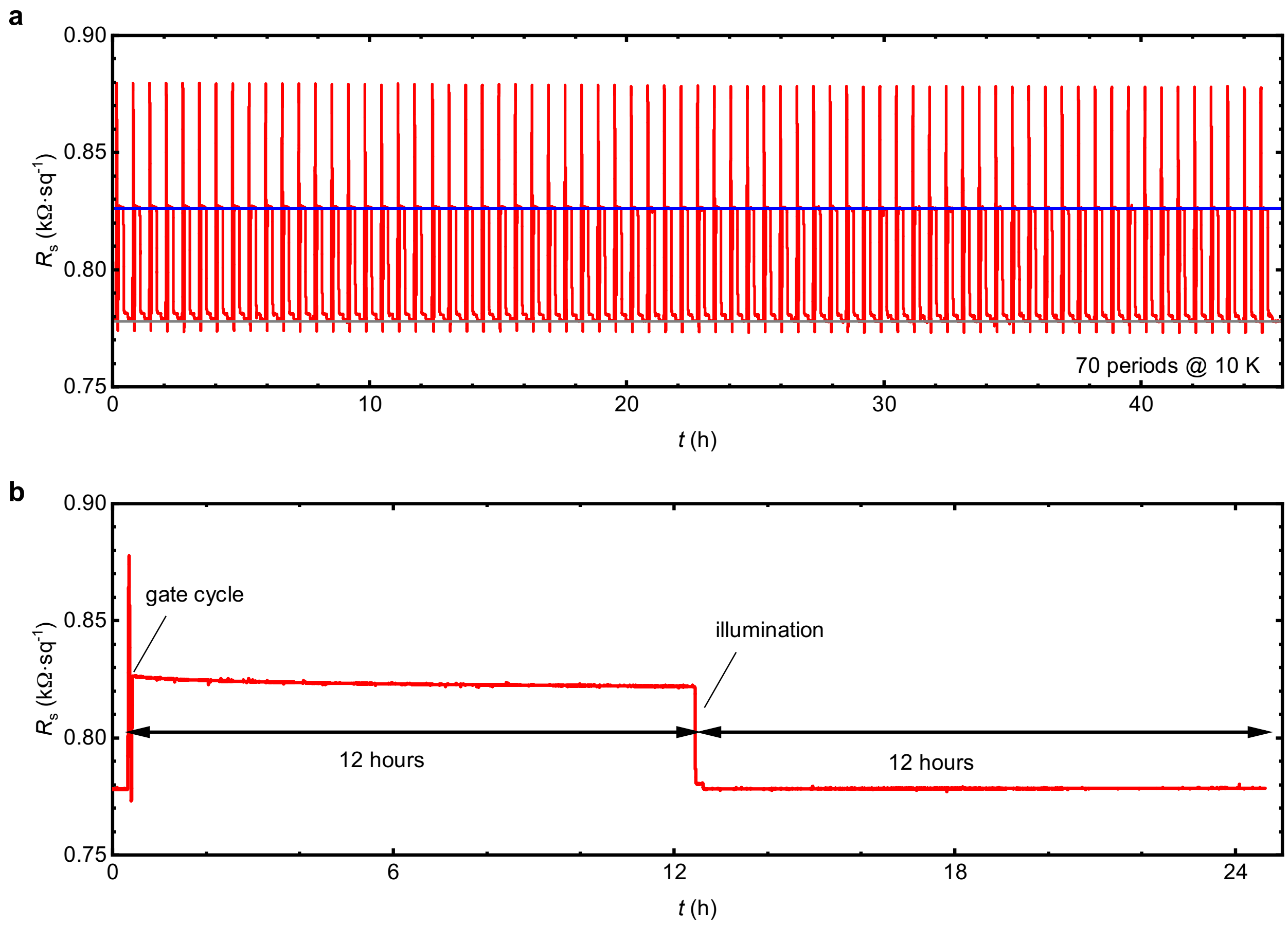

**Supplementary Figure 10. Reproducibility and non-volatility tests. a,** $R_s$ variation during 70 periods of gate cycles and optical illuminations. The blue and green lines track the value of $R_s$ after each gate cycle and illumination, respectively. **b,** $R_s$ variations for 12 hours after a gate cycle and an illumination period. No obvious decay is observed.

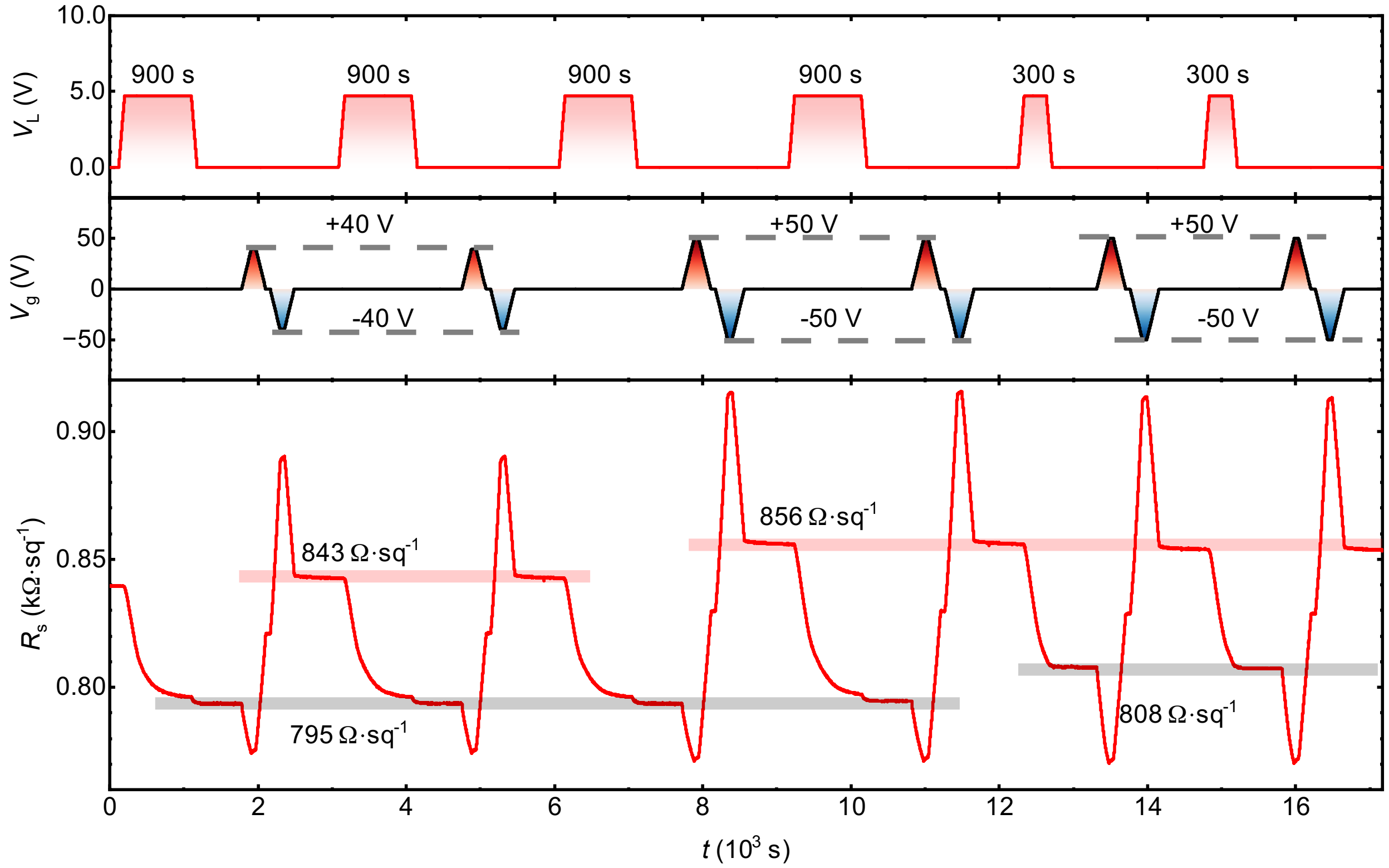

**Supplementary Figure 11.** $\boldsymbol{R_s}$ **variation during a sequence of illuminations and gate cycles at 10 K.** The top and middle panels display the temporal-series of illuminations (represented by the voltage applied on the LED, $V_L$) and gate cycles (represented by the gate voltage $V_g$). The bottom panel presents the variation of $R_s$. The grey bars mark the $R_s$ baselines after illumination, and the red bars mark its elevated values after gate cycles.

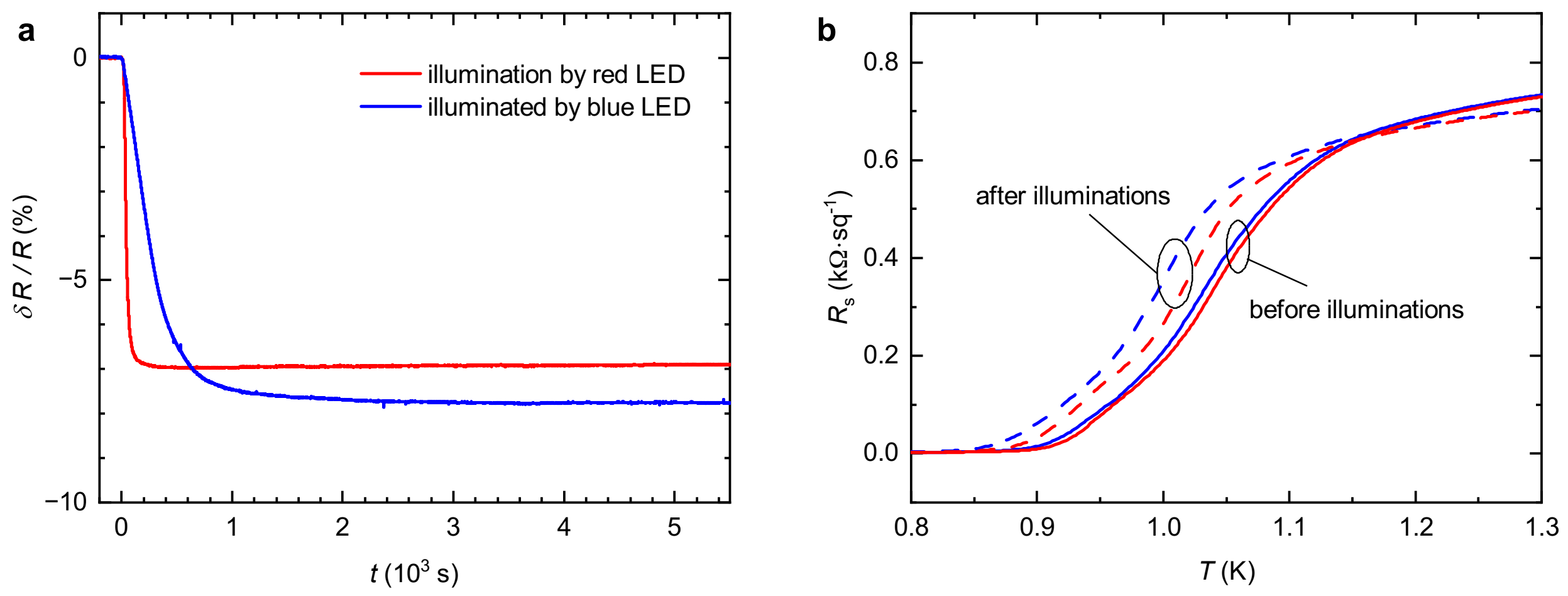

**Supplementary Figure 12. Effect of illuminations by red and blue lights. a,** Temporal evolutions of $R_s$ after illuminated by red and blue LEDs. The blue light leads to a larger but slower resistance drop than the red light. **b,** Shifts of superconducting transition by illuminations of red (by red curves) and blue lights (by blue curves). The solid curves represent the transitions before the illuminations, and the dashed curves represent the transitions after the illuminations.

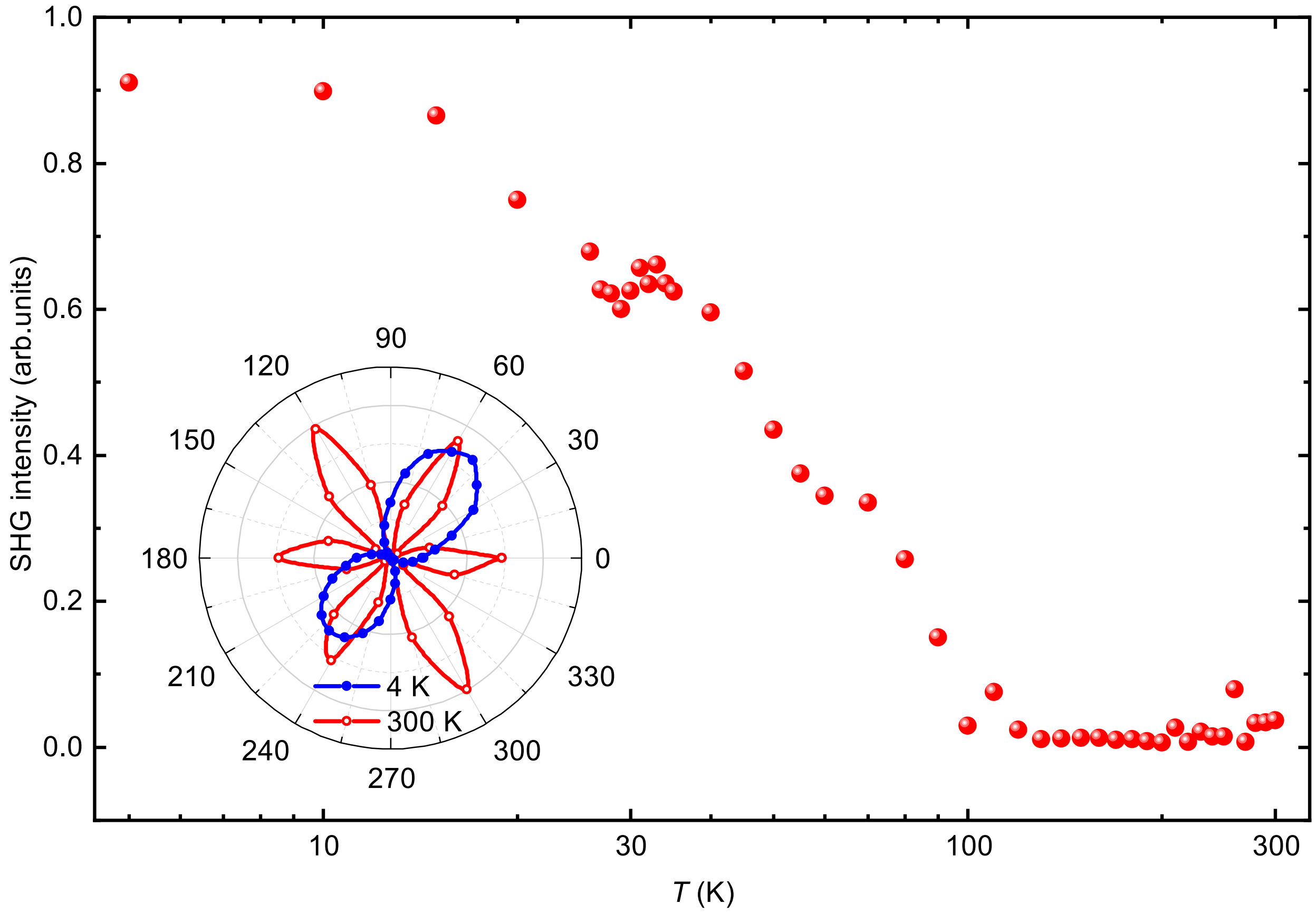

**Supplementary Figure 13. Intensity of second harmonic generation (SHG) as a function of temperature.** The inset displays the angular dependence of SHG intensity measured at 300 K and 4 K.

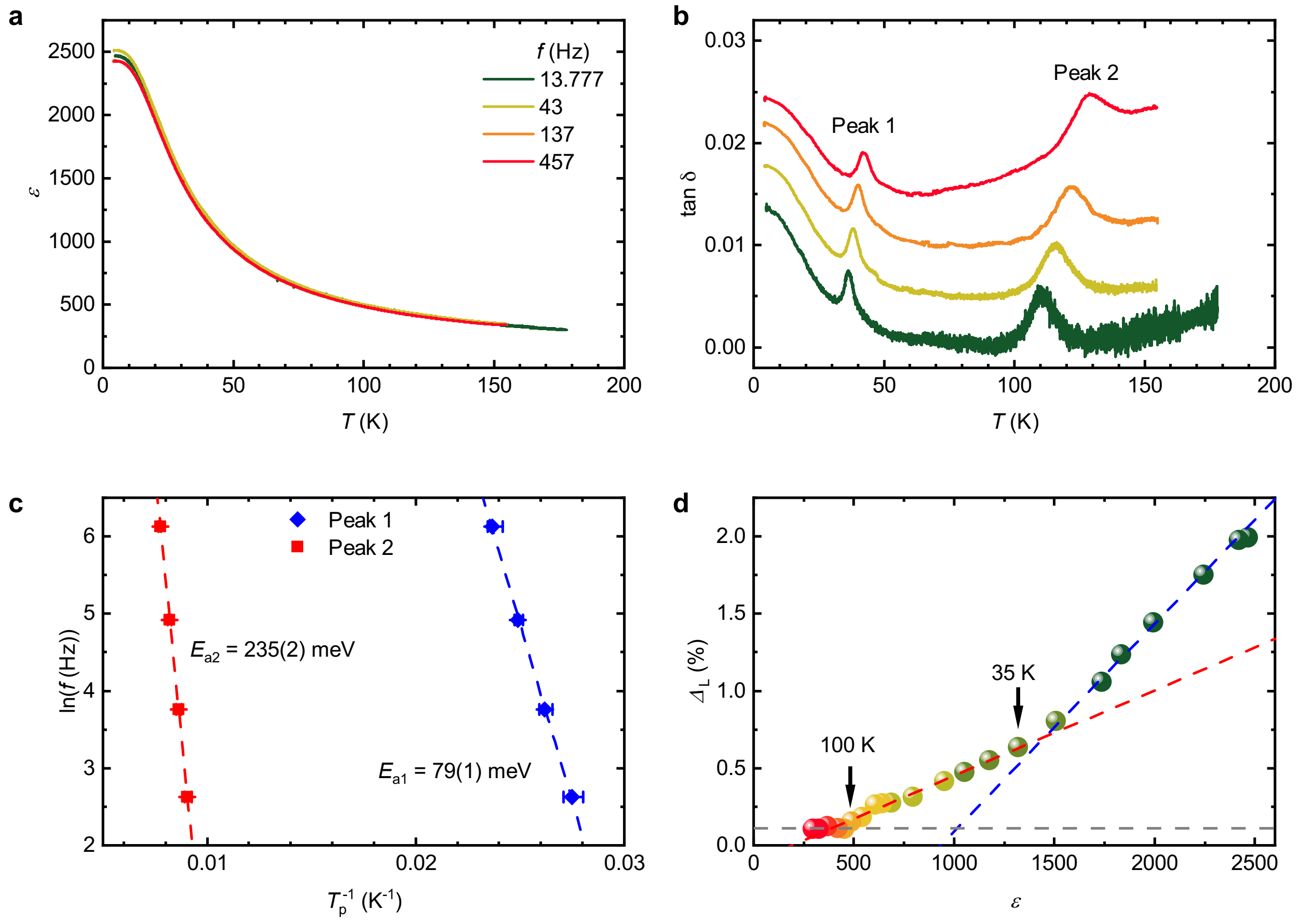

**Supplementary Figure 14. Dielectric responses for S2. a,** Temperature dependence of dielectric constant ($\varepsilon$) measured at multiple frequencies. **b,** Temperature dependence of dielectric tangent loss ($\tan\delta$) measured at multiple frequencies. **c,** Arrhenius analyses of the peak temperatures for Peak 1 and Peak 2. The error bars are determined by the temperatures at which the magnitude of $\tan\delta$ deviates below its peak value by one standard deviation. The red and blue dashed lines represent the Arrhenius fittings of the data points. **d,** Correlation between the hysteretic resistance loop size ($\Delta_{\mathrm{L}}$) and $\varepsilon$. $\Delta_{\mathrm{L}}$ is extracted from the $R_{\mathrm{s}}$ loops shown in Supplementary Figure 7. The red and blue dashed lines depict the linear dependence below 100 K and 35 K, respectively. The horizontal grey dashed line represents the constant $\Delta_{\mathrm{L}}$ beyond 100 K.

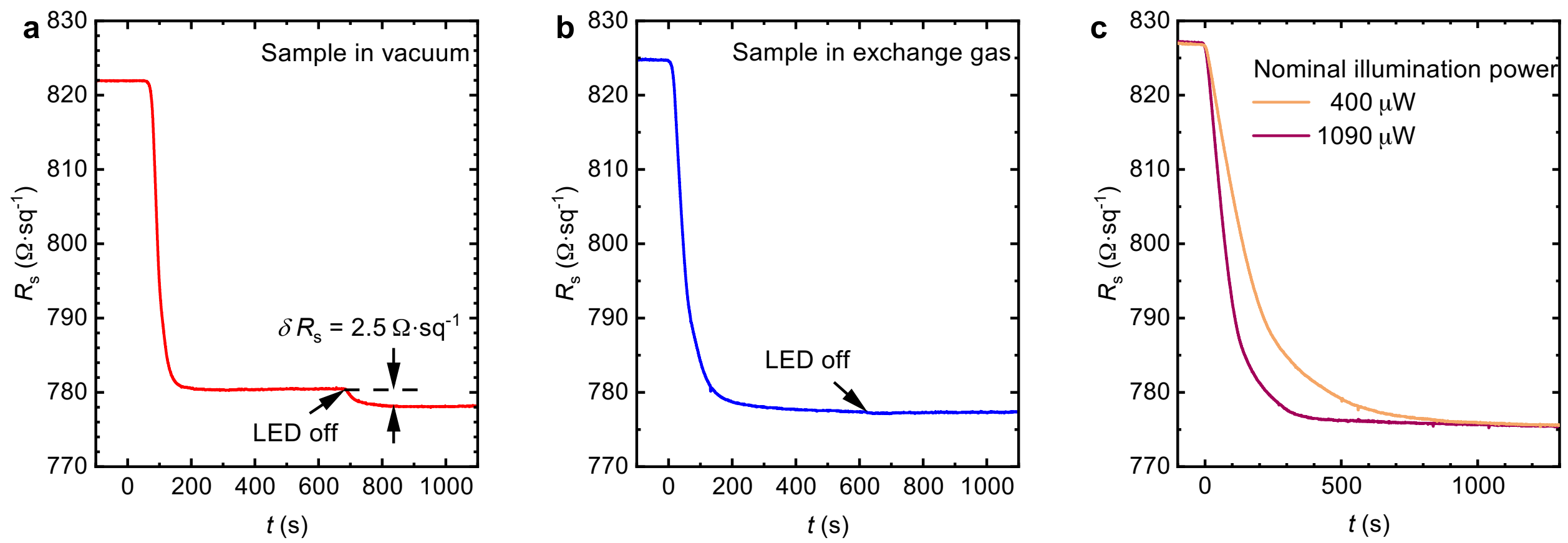

**Supplementary Figure 15. Effect of illuminations on $\boldsymbol{R_s}$ at various conditions. a,** $R_s$ variation when the sample is illuminated at 10 K in vacuum. **b,** $R_s$ variation when the sample is illuminated at 10 K in exchange gas of helium. **c,** Comparison of $R_s$ as the sample is illuminated by different powers.

**Supplementary Table 1. Specifics of $AlO_x$/ $KTaO_3$(KTO) samples.** Values of carrier density ($n$) and mobility ($\mu$) are determined by Hall measurements at normal state after gate cycles.

| Sample | Thickness of Al deposition before oxidization (nm) | $n$ ($10^{14}$ $cm^{-2}$) | $\mu$ ($cm^2 \cdot V^{-1} \cdot s^{-1}$) |
|---|---|---|---|
| S1 | 2.5 | 1.06 | 77 |
| S2 | 1.5 | 0.77 | 56 |
| S3 | 2.0 | 0.95 | 33 |